\documentclass[aps,prd,reprint,a4paper,showpacs,nofootinbib,superscriptaddress]{revtex4-2}

\usepackage{bbm}
\usepackage{amsmath}
\usepackage{graphicx}
\usepackage{float}
\usepackage{amssymb}%,ulem
\usepackage{subfigure}
\usepackage{amssymb}
\usepackage{hyperref}
\usepackage{epstopdf}

\usepackage[utf8]{inputenc}
\hypersetup{
    colorlinks=true,
    linkcolor=red,
    citecolor=blue,
}
\usepackage{color}
\usepackage[T1]{fontenc}
\usepackage{txfonts}

\begin{document}

\title{On an alternative mechanism for the black hole echoes}
\author{Hang Liu}
\email{hangliu@sjtu.edu.cn}
\affiliation{Center for Gravitation and Cosmology, College of Physical Science and Technology, Yangzhou University, Yangzhou 225009, China}
\affiliation{School of Physics and Astronomy, Shanghai Jiao Tong University, Shanghai 200240, China}

\author{Wei-Liang Qian}
\email{wlqian@usp.br}
\affiliation{Escola de Engenharia de Lorena, Universidade de S\~ao Paulo, 12602-810, Lorena, SP, Brazil}
\affiliation{Center for Gravitation and Cosmology, College of Physical Science and Technology, Yangzhou University, Yangzhou 225009, China}
\affiliation{Institute for theoretical physics and cosmology, Zhejiang University of Technology, 310032, Hangzhou, China}

\author{Yunqi Liu}
\affiliation{Center for Gravitation and Cosmology, College of Physical Science and Technology, Yangzhou University, Yangzhou 225009, China}

\author{Jian-Pin Wu}
\affiliation{Center for Gravitation and Cosmology, College of Physical Science and Technology, Yangzhou University, Yangzhou 225009, China}

\author{Bin Wang}
\email{wang_b@sjtu.edu.cn}
\affiliation{Center for Gravitation and Cosmology, College of Physical Science and Technology, Yangzhou University, Yangzhou 225009, China}
\affiliation{School of Aeronautics and Astronautics, Shanghai Jiao Tong University, Shanghai 200240, China}

\author{Rui-Hong Yue}
\affiliation{Center for Gravitation and Cosmology, College of Physical Science and Technology, Yangzhou University, Yangzhou 225009, China}

\begin{abstract}
Gravitational wave echoes from the black holes have been suggested as a crucial observable to probe the spacetime in the vicinity of the horizon.
In particular, it was speculated that the echoes are closely connected with specific characteristics of the exotic compact objects, and moreover, possibly provide an access to the quantum nature of gravity.
Recently, it was shown that the discontinuity in the black hole metric substantially modifies the asymptotical behavior of quasinormal frequencies.
In the present study, we proceed further and argue that a discontinuity planted into the metric furnishes an alternative mechanism for the black hole echoes.
Physically, the latter may correspond to an uneven matter distribution inside the surrounding halo.
To demonstrate the results, we first numerically investigate the temporal evolution of the scalar perturbations around a black hole that possesses a nonsmooth effective potential.
It is shown that the phenomenon persists even though the discontinuity can be located further away from the horizon with rather insignificant strength.
Besides, we show that the echoes in the present model can be derived analytically based on the modified pole structure of the associated Green function.
The asymptotical properties of the quasinormal mode spectrum and the echoes are found to be closely connected, as both features can be attributed to the same origin.
In particular, the period of the echoes in the time domain $T$ is shown to be related to the asymptotic spacing between successive poles along the real axis in the frequency domain $\Delta(\Re\omega)$, by a simple relation $\lim\limits_{\Re\omega\to+\infty}\Delta(\Re\omega) = 2\pi/T$.
Moreover, we discuss possible distinguishment between different echo mechanisms.
The potential astrophysical implications of the present findings are also addressed.
\end{abstract}

%\date{\today}
\date{July 07th, 2021}

\maketitle

\section{Introduction}

The unprecedented detections of the gravitational waves (GWs) by the LIGO and Virgo Collaboration~\cite{TheLIGOScientific:2016agk} directly confirmed the last remaining piece of predictions by Einstein's general relativity (GR).
It was designated as a landmark for the advent of GW astronomy.
On the experimental side, inspired by the success of the ground-based facilities, several ongoing spaceborne GW detection programs are under active development.
The latter include the LISA~\cite{Audley:2017drz}, TaiJi~\cite{Hu:2017mde}, and TianQin~\cite{Luo:2015ght} projects, which aim to scrutinize, specifically, the GWs at low frequencies.
On the theoretical side, the significance of GW detection resides in the possibility of probing the Universe through gravitational interactions in a more extensive fashion.
In particular, GW opens up new avenues to furnish crucial information in addition to those obtained by traditional approaches primarily based upon electromagnetic radiations.
Subsequently, a variety of topics, essentially associated with the foundation of modern physics, are renovated accordingly.
Indeed, the inauguration of a novel era is heralded with further quantitative investigations of dark matter~\cite{Obata:2018vvr,Pierce:2018xmy,Liu:2018icu,Nagano:2019rbw,Manley:2020mjq,Stadnik:2014tta,Grote:2019uvn}
and dark energy~\cite{Weiner:2020sxn,Garoffolo:2020vtd,Singh:2020nna,Noller:2020afd},
as well as feasible discrimination between different theories of modified gravity~\cite{Nunes:2020rmr,Mastrogiovanni:2020gua},
which are partly facilitated by the more accurate estimation of the cosmological parameter using the standard GW siren~\cite{Zhao:2019gyk,Zhang:2019mdf,Zhang:2019loq,Zhang:2019ylr,Wang:2019tto,Jin:2020hmc}.

Besides these promising possibilities, the studies of the black hole echo also have attracted much attention recently~\cite{Cardoso:2016rao,Cardoso:2016oxy,Foit:2016uxn,Bueno:2017hyj,agr-qnm-echoes-14,agr-qnm-echoes-15,Burgess:2018pmm,Konoplya:2018yrp, Testa:2018bzd,Wang:2018mlp,Cardoso:2019apo,Ghersi:2019trn,Wang:2019szm,Liu:2020qia,agr-qnm-echoes-18}.
The notion of GW echoes was proposed in close connection with the GW ringdown signals, targeting the physics in the vicinity of the black holes and their exotic compact alternatives.
The latter is referred to in the literature as horizonless exotic compact objects (ECOs)~\cite{Cardoso:2019rvt}.
The ringdown stage of the perturbations around such objects may possess a pattern similar to that of the black holes, but followed by subsequential pulses with unique characteristics, known as the {\it echoes}~\cite{Cardoso:2016oxy}.
The relevant physical systems involve wormholes~\cite{Cardoso:2016rao,Bueno:2017hyj},
gravastars~\cite{Visser:2003ge}, boson stars~\cite{Schunck:2003kk},
anisotropic stars~\cite{Carloni:2017bck},
dark stars~\cite{Baccetti:2018qrp,Giddings:1992hh},
fuzzballs~\cite{Lunin:2002qf,Mathur:2012jk}, and firewalls~\cite{Mathur:2012jk,Almheiri:2012rt}.

On the one hand, as it was argued~\cite{Konoplya:2011qq}, the black hole quasinormal modes are mostly governed by the spacetime in the vicinity of the horizon.
The echoes, on the other hand, are understood to be related to the properties of the light ring and the surface, in the proximity to the would-be horizon, of the compact object~\cite{Cardoso:2016rao}.
In this context, the feasibility regarding the measurement of GW echoes is particularly interesting as no existing experiment has been able to explicitly probe the relevant region of the spacetime~\cite{agr-bh-exp-01,agr-bh-exp-05,agr-bh-exp-06}.
To be more specific, the GW echoes may play an essential role as a crucial agency to access the distinctive properties of the ECOs or might even offer us some hints regarding quantum gravity~\cite{Cardoso:2016oxy, Foit:2016uxn}.

In the literature, it has been speculated that the presence of the GW echoes is due to the back-and-forth bouncing of the wave-packet trapped in the potential-well bounded between two peaks of the effective potential.
The specific shape of the effective potential in question might be formed by the centrifugal barrier or the quantum corrections to the conventional black hole horizons~\cite{Cardoso:2016rao,Cardoso:2016oxy,Foit:2016uxn,Bueno:2017hyj,agr-qnm-echoes-15,Burgess:2018pmm,Konoplya:2018yrp, Testa:2018bzd,Wang:2018mlp,Cardoso:2019apo,Wang:2019rcf,Ghersi:2019trn,Wang:2019szm}.
For instance, one peak of the desired potential can be constituted by the maximum of the Regge-Wheeler potential, and moreover, a second peak is introduced in accordance with appropriate physical considerations.

From another perspective in terms of the Green function, however, it was pointed out that the distinct echo feature can be derived by analyzing the resonant frequencies of the ECO~\cite{agr-qnm-echoes-15}.
In particular, there is a close relationship between the ECO quasinormal modes and those of the associated black hole.

Recently, it was shown that discontinuity appreciably affects the asymptotic properties of the black hole quasinormal frequencies~\cite{agr-qnm-22, agr-qnm-30, Qian:2020cnz} as well as the shadow~\cite{Qian:2021qow}.
In particular, it was pointed out in Ref.~\cite{Qian:2020cnz} that even a minor discontinuity in the effective potential may lead to a significant modification to the behavior of the quasinormal modes of higher overtone numbers.
The origin of the results was partly attributed to the fact that the WKB approximation becomes invalid at the point of discontinuity.
Subsequently, the asymptotics of the Wronskian, as well as the location of the pertinent poles, is modified.
On a rather different ground, the discontinuity was found to give rise to cuspy and fractured black hole shadows~\cite{Qian:2021qow}.
The resultant distortion to the shadow edge can be understood in terms of the feasibility of the Maxwell construction.
In both cases, the discontinuity in the effective potential can be implemented by introducing an infinitesimally thin dark matter shell wrapped around a (rotating) black hole.

As will be discussed further, one may argue that the above spacetime configuration is meaningful from an astrophysical point of view.
This is primarily because the astrophysical objects live in a non-vacuum environment.
Possible relevant scenarios include the pulsating stars, accreting disks, active galactic nuclei, clouds of normal and/or phantom matter, and objects related to the concept of ``dirty black hole''~\cite{Visser:1992qh,Visser:1993qa,Visser:1993nu,Krauss:1996rg,Macedo:2015ikq,Babichev:2004yx}.
As a result, the GWs emanated from the vicinity of the black hole are not free from the influence of the environments, and as it turns out, they are subject to nontrivial modifications.
To a reasonable degree, by placing a matter shell around the black hole, one devises a toy model to mimic the complicated contents regarding the matter distribution outside the black hole.
Therefore, it is natural to ask to what extent such a simplified configuration will affect the propagation of the GWs~\cite{Barausse:2014tra}, and furthermore, whether the echoes might be created.
As discussed above, the GW echoes are shown to be associated with the properties of the Green function, whose poles define the resonance frequencies of the system.
Therefore, it seems intriguing by further taking into consideration that the asymptotic properties of the latter are shown to be rather sensible to discontinuity in the effective potential~\cite{Qian:2020cnz}.
In this context, the resultant modification of the pole structure of the Green function might lead to sizable modification of the propagation of the GWs.
The above considerations serve as our primary motivation.

The present study involves an attempt to explore the effect of discontinuity on the propagation of GWs.
To be specific, we report that the echoes of the GW wave-packet can be observed when a discontinuity is planted in the effective potential.
It is somewhat less intuitive that for the present scenario, the echoes are formed between the maximum of the Regge-Wheeler potential and the point of the discontinuity.
The latter carries the role of the second maximum in the effective potential, which turns out to be feasible even it is furnished by a downward step.
It is shown that the echoes persist even though the size of the step is rather insignificant and located further away from the horizon.
Moreover, we proceed further to explore its physical origin in terms of the pole structure of the Green function in the frequency domain.
By analyzing the modified asymptotics of the quasinormal frequencies, we demonstrate its underlying connection to the echoes observed numerically.
We explicitly derive a simple relation between the period of the echoes in the time domain and the asymptotic spacing between the poles of the Green function in the frequency domain.
The relation is first derived for a simplified case, then it is demonstrated that it remains valid on general grounds.

The remainder of the present work is organized as follows.
In Sec.~\ref{section2}, we briefly review the problem of the scalar perturbations in a static black hole metric.
In Sec.~\ref{section3}, the numerical method employed in our study is presented.
Subsequently, the effects of the discontinuity in the potential on the GWs propagation and the resultant echoes are investigated numerically.
In particular, we devise and explore two types of metric setups regarding different matter distributions.
In Sec.~\ref{section4}, analytic analyses are carried out for the pole structure of the Green function.
The resultant modification to the quasinormal modes of the original metric in the frequency domain, as well as their connection to the echoes in the time domain, are established.
The last section is devoted to further discussions and concluding remarks.

\section{The model} \label{section2}

In this section, we briefly outline the physical problem by presenting the devised spacetime metric together with the relevant equation of motion and its boundary conditions.
As discussed in the last section, we propose to introduce a small discontinuity into an existing black hole metric at a location significantly far away from the horizon.
Here, our main goal is to investigate the resultant effect on the propagation of the GWs, and in particular, the possible emergence of echoes deviating from the standard quasinormal oscillations of the original black hole metric.

For simplicity, a static black hole metric in $3+1$ spacetime with spherical symmetry $g_{\mu\nu}$ is considered,
\begin{equation}
ds^2=-A(r)dt^2+B(r)dr^2+r^2(d\theta^2+\sin^2\theta d\phi^2) .
\end{equation}
Also, we will only focus on the perturbations of a scalar field, as it turns out to suffice to capture the underlying physics.
The equation of motion of a massless scalar field $\Phi$ reads
\begin{eqnarray}
\frac{1}{\sqrt{-g}}\partial_{\mu}(\sqrt{-g}g^{\mu\nu}\partial_{\nu}\Phi(t,r,\theta,\phi))=0.\label{eq5}
\end{eqnarray}
Using the method of separation of variables for $\Phi(t,r,\theta,\phi)$, namely,
\begin{equation}
\Phi(t,r,\theta,\phi)=\sum_{l,m}\frac{\Psi(t,r)}{r}Y_{lm}(\theta,\phi)\label{eq4},
\end{equation}
where $Y_{lm}(\theta,\phi)$ is the spherical harmonics and $l$ and $m$ stand for the angular and azimuthal number, respectively.

Substituting Eq. \eqref{eq4} into Eq. \eqref{eq5}, one obtains the following radial equation
\begin{equation}
\begin{split}
-\partial_t^2\Psi(t,r)&+\frac{A}{B}\partial_r^2\Psi(t,r)+\frac{BA'-AB'}{2B^2}\partial_r\Psi(t,r)\\
&+\frac{A(rB'-2l(l+1)B^2)-rBA'}{2r^2B^2}\Psi(t,r)=0,\label{eq6}
\end{split}
\end{equation}
where a prime denotes a derivative with respect to areal radius $r$.

By making use of the tortoise coordinate $x$ defined by
\begin{equation}
dx=\sqrt{\frac{B(r)}{A(r)}}dr,
\end{equation}
Eq.~\eqref{eq6} can be rewritten as
\begin{equation}
-\frac{\partial^2\Psi(t, r)}{\partial t^2}+\frac{\partial^2\Psi(t,r)}{\partial x^2}-V(r)\Psi(t, r)=0 , \label{masterEqT}
\end{equation}
where the effective potential $V(r)$ is given by
\begin{equation}
V(r)=A(r)\frac{l(l+1)}{r^2}+\frac{1}{2r}\frac{d}{dr}\frac{A(r)}{B(r)}.\label{eqsc}
\end{equation}

For the remainder of the present paper, we restrict our consideration to the following metric
\begin{equation}
A(r)=\frac{1}{B(r)}\equiv f(r)=1-\frac{2m(r)}{r} ,
\end{equation}
where the mass function $m(r)$ gives the total mass contained within the radius $r$.
For the purpose of the present study, the form of the mass function will be tuned to indicate a thin layer of mass shell wrapped around a black hole at the center.
The specific form of the function $m(r)$ will be discussed in the next section.
Accordingly, the effective potential Eq.~\eqref{eqsc} simplifies to read
\begin{equation}\label{eq7}
V(r)=f(r)\left(\frac{l(l+1)}{r^2}+\frac{f'(r)}{r}\right) ,
\end{equation}
which readily falls back to the Regge-Wheeler potential when $m(r)=M_0$
\begin{equation}\label{eqRW}
V_{\mathrm{RW}}(r)=\left(1-\frac{2M_0}{r}\right)\left(\frac{l(l+1)}{r^2}+\frac{2M_0}{r^3}\right) .
\end{equation}

The black hole quasinormal modes is determined by solving the eigen-value problem defined by Eq.~\eqref{masterEqT} in the frequency domain
\begin{equation}
\frac{d^2\Psi(\omega, x)}{dx^2}+(\omega_{ln}^2-V(r))\Psi(\omega, x) = 0 , \label{eq2}
\end{equation}
for $\Psi$ with the following boundary conditions for asymptotically flat spacetimes
\begin{equation}
\Psi \sim
\begin{cases}
   e^{-i\omega_{ln} x}, &  x \to -\infty, \\
   e^{+i\omega_{ln} x}, &  x \to +\infty,
\end{cases}
\label{master_bc0}
\end{equation}
which indicate ingoing wave at the horizon and outoging wave at infinity.
Here, the eigen value $\omega_{ln}$ is known as the quasinormal frequency, it is usually a complex number due to the dissipative nature of Eq.~\eqref{master_bc0}.
The subscript $l$ denotes the angular number and $n$ represents the overtone number.

\section{Numerical results of the echoes}\label{section3}

In this section, the temporal evolution of the scalar perturbations in the metric presented in the previous section is studied numerically.
We first discuss the employed numerical scheme and then present the obtained results.

\subsection{The finite difference method}

The time-domain profiles of the scalar field can be obtained by directly integrating the partial differential equation Eq.~\eqref{masterEqT} in terms of the coordinates $(t, x)$.
To this end, we descretize the coordinates and denote $\Psi(t,x)=\Psi(i\Delta t, j\Delta x)=\Psi_{i,j}$, $V(r(x))=V(j\Delta x)=V_{j}$, such that Eq.~\eqref{masterEqT} is rewritten as
\begin{equation}
\begin{split}
-&\frac{(\Psi_{i+1,j}-2\Psi_{i,j}+\Psi_{i-1,j})}{\Delta t^2}+\frac{(\Psi_{i,j+1}-2\Psi_{i,j}+\Psi_{i,j-1})}{\Delta x^2}\\
&-V_{j}\Psi_{i,j}+O(\Delta t^2)+O(\Delta x^2)=0.
\end{split}
\end{equation}
to the second order~\cite{Zhu:2014sya}.
Thus, the evolution of field on the grid sites is governed by
\begin{equation}
\Psi_{i+1,j}=-\Psi_{i-1,j}+\frac{\Delta t^2}{\Delta x^2}(\Psi_{i,j+1}+\Psi_{i,j-1})\\
+(2-2\frac{\Delta t^2}{\Delta x^2}-\Delta t^2V_{j})\Psi_{i,j}.
\end{equation}

By considering an initial distribution that has compact support in tortoise coordinate, the boundary condition is effectively free as long as the signal has not reached the bound of the grids.
In this sense, the boundary condition defined in Eq.~\eqref{master_bc0} is irrelevant.
In practice we adopt the following initial Gaussian distribution $\Psi(t=0,x)=\mathrm{exp}[-\frac{(x-a)^2}{2b^2}]$ and $\Psi(t<0,x)=0$.
We take $b=3$, and the parameter $a$ will be chosen accordingly for different cases.
The von Neumann stability condition can be fullfilled by confining the calculations to an appropriate spatial region while keeping a small ratio $\Delta t/\Delta x < 1$.
Accordingly, in our calcultions, we take $\Delta t=0.1$ and $\Delta x=0.2$.

In the following two subsections, we proceed to present numerically the occurrence of echoes in our model.
We explore metrics with two types of discontinuity.
In the next subsection, we first consider the cases where the effective potentials possess jump discontinuity.
Then, we turn to investigate the cases where the effective potentials are continuous functions where the discontinuity only appears in the derivative.

\subsection{Echoes in potential function with $\mathcal{C}^0$ discontinuity}

\begin{figure}
\centering
\includegraphics[height=2in,width=3in]{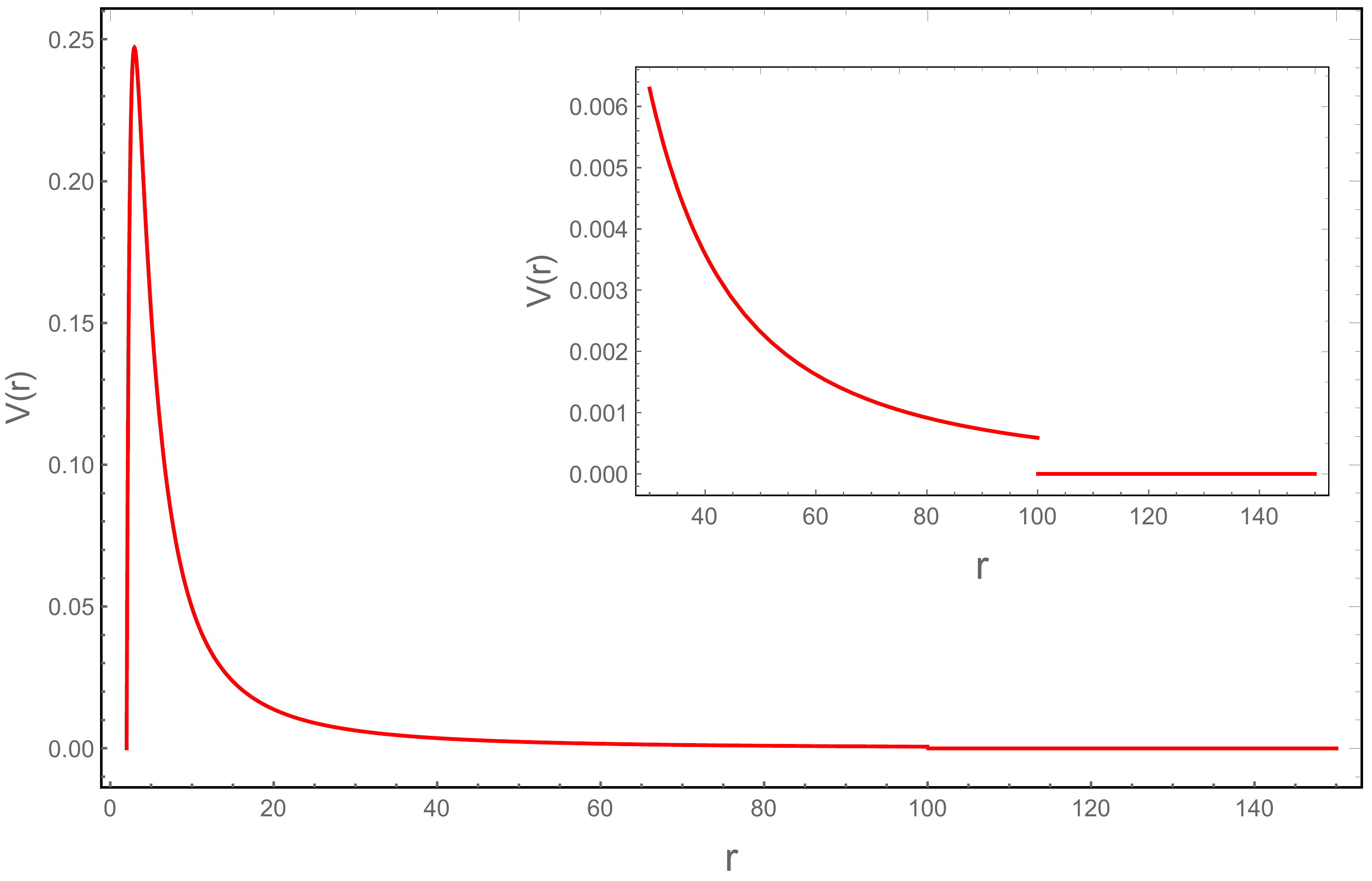}
\caption{The effetive potential given in Eq.~\eqref{eq1}, where we take $l=2$, $M_0=1$, and $r_s=100$.}\label{fig1}
\end{figure}

Now we are in a position to present the numerical results.
In the present and following subsections, by investigating the temporal evolution of the perturbation in various metrics with discontinuity, we demonstrate the robustness of the echoes in the proposed picture.
Moreover, we argue that the resulting phenomenon is independent of the strength and location of the discontinuity.
In other words, the echoes persist even though the discontinuity can be located further away from the horizon with rather insignificant strength.

We start by the following spacetime where the power-law tail of the Regge-Wheeler potential Eq.~\eqref{eqRW} is cut off at $r=r_s$, namely,
\begin{equation}\label{eq1}
V_1(r) =
\begin{cases}
 \left(1-\frac{2M_0}{r}\right)\left(\frac{l(l+1)}{r^2}+\frac{2M_0}{r^3}\right)   &  r\leq r_s, \\
 0 &  r>r_s ,
\end{cases}
\end{equation}
which introduces a jump discontinuity at $r=r_s$, as shown in Fig.~\ref{fig1}.
It is worth noting that the ``cut'' in the effective potential is visually insignificant, which also holds true for the other metrics investigated in this section.

\begin{figure}
\centering
\includegraphics[height=2in,width=3in]{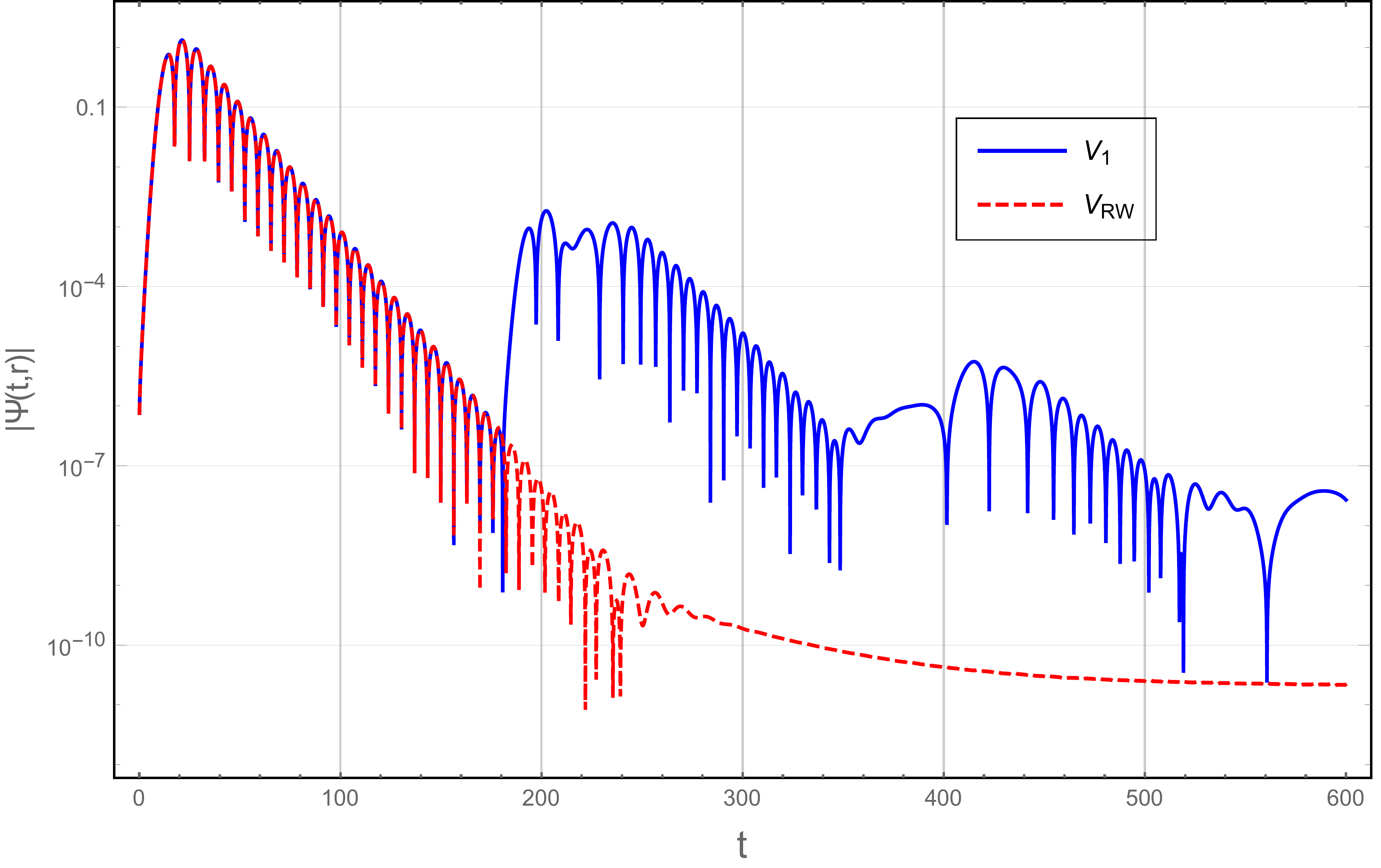}
\caption{The time-domain profile of scalar field perturbations for the effective potential $V_1$ defined by Eq.~\eqref{eq1} (blue solid curve), compared against with that for the Regge-Wheeler potential $V_{\mathrm{RW}}$ (red dashed curve).
The calculations are carried out by taking $l=2$, $M_0=1$, and $r_s=100$.}\label{fig2}
\end{figure}

The corresponding temporal evolution for the scalar perturbations is obtained and presented in Fig.~\ref{fig2}.
For comparison, we show the time-domain profiles of the scalar field perturbations for both the Regge-Wheeler potential and that for $V_1$ defined by Eq.~\eqref{eq1}.
Although the initial quasinormal ringings are largely identical in both cases, it is evident that the modification of the Regge-Wheeler led to a significant change of the waveform for $t\gtrsim 180$.
On the one hand, for the Regge-Wheeler, the profile features the quasinormal ringing followed by a late-time tail.
For the modified potential $V_1$, on the other hand, the waveform consists of successive echoes.
The late-time tail has not been observed in this case and is not expected to, as a result of the truncation in the potential~\cite{agr-qnm-tail-06}.
One also observes that the delay between successive echoes is mostly constant.
The interval is in accordance with the time for the wave to travel and bounced back from the ``cut''.
Also, the identical initial waveforms observed for $t\in[0,180]$ can be attributed to the causality.

\begin{figure}
\centering
\includegraphics[height=2in,width=3in]{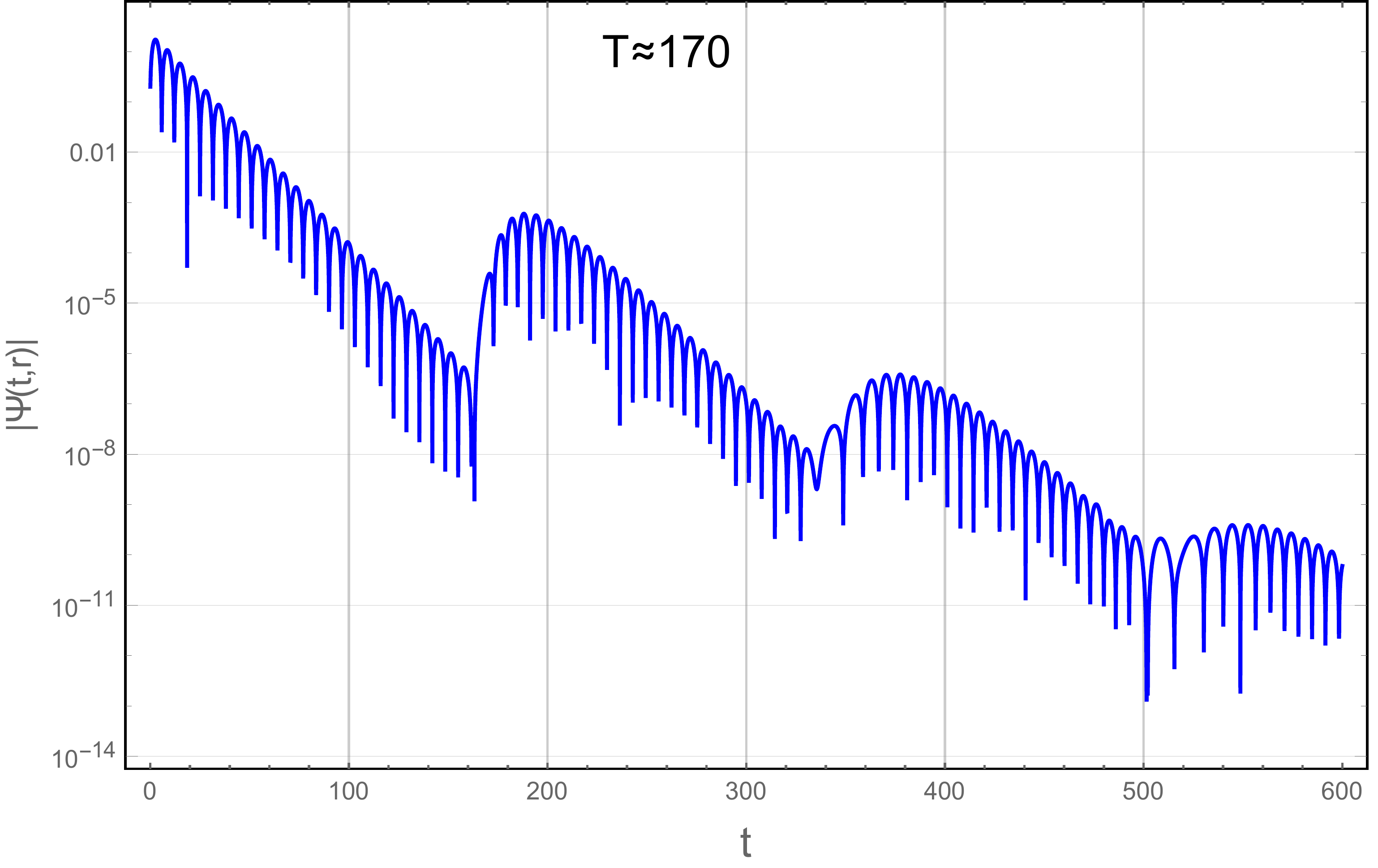}
\includegraphics[height=2in,width=3in]{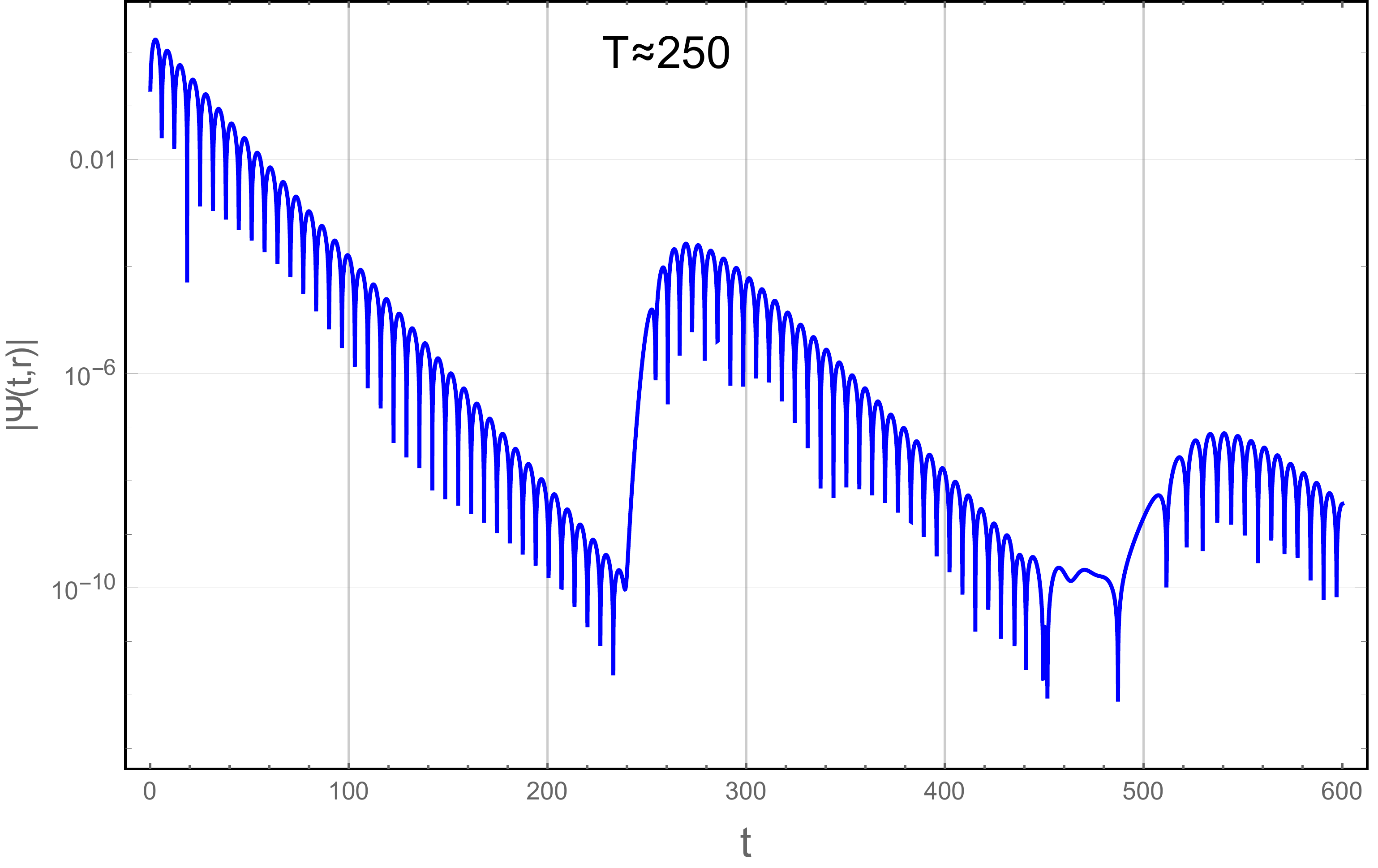}
\caption{The time-domain profile of scalar field perturbations for the effective potential $V_1$ defined by Eq.~\eqref{eq1}.
The calculations are carried out by taking $l=2$, $M_0=1$, and for $r_s=80$ and $r_s=120$, respectively.}\label{fig3}
\end{figure}

To support the above observations, in Fig.~\ref{fig3}, we recalculate the time-domain profiles of the scalar perturbations, while using slightly different forms of the potential $V_1$.
To be specific, the calculations are carried out adopting the same initial conditions as those in Fig.~\ref{fig1}, but using different paramters $r_s=80$ and $r_s=120$ for the potential.
It is found that the echoes persist while the initial quasinormal ringings largely remain identical before the echoes take place.
For the same reason discussed before, the late-time tail is not observed.
Also, the strength of the echoes seems not sensitive to the location of the discontinuity, however, the echoes in the two cases are featured by different periods of recurrence.
To be specific, the resultant intervals between successive echoes are found to be, respectively, $T\sim 170$ and $T\sim 250$ as indicated in the plots.
These values confirmed our above intuition that the echoes are formed due to the reflection of waves between the maximum of the Regge-Wheeler potential and the planted discontinuity.
We note that the latter is not a peak of the potential but rather a ``dent''.
Indeed, it is readily verified numerically that $T=\Delta x =2 (x_s-x_{\mathrm{ph}}) \sim 2 x_s$, where $x_s=x(r_s)$ is the tortoise coordinate of the discontinuity and $x_{\mathrm{ph}} = x(r_{\mathrm{ph}})$ where $r_{\mathrm{ph}}=3 M_0$ is the radius of the photon sphere.

Although the effective potential given in Eq.~\eqref{eq1} is simple in its apparent form, it corresponds to a spherically symmetric black hole surrounded by a matter distribution for $r\ge r_s$, namely,
\begin{equation}\label{eqm1}
m_1(r) =
\begin{cases}
M_0,  &  r\leq r_s, \\
\frac{1}{2}\left[l(l+1)\ln\left(\frac{r}{r_s}\right)+\frac{2M_0}{r_s}\right]r, &  r>r_s,
\end{cases}
\end{equation}
which is continuous at $r=r_s$.

\begin{figure}
\centering
\includegraphics[height=2in,width=3in]{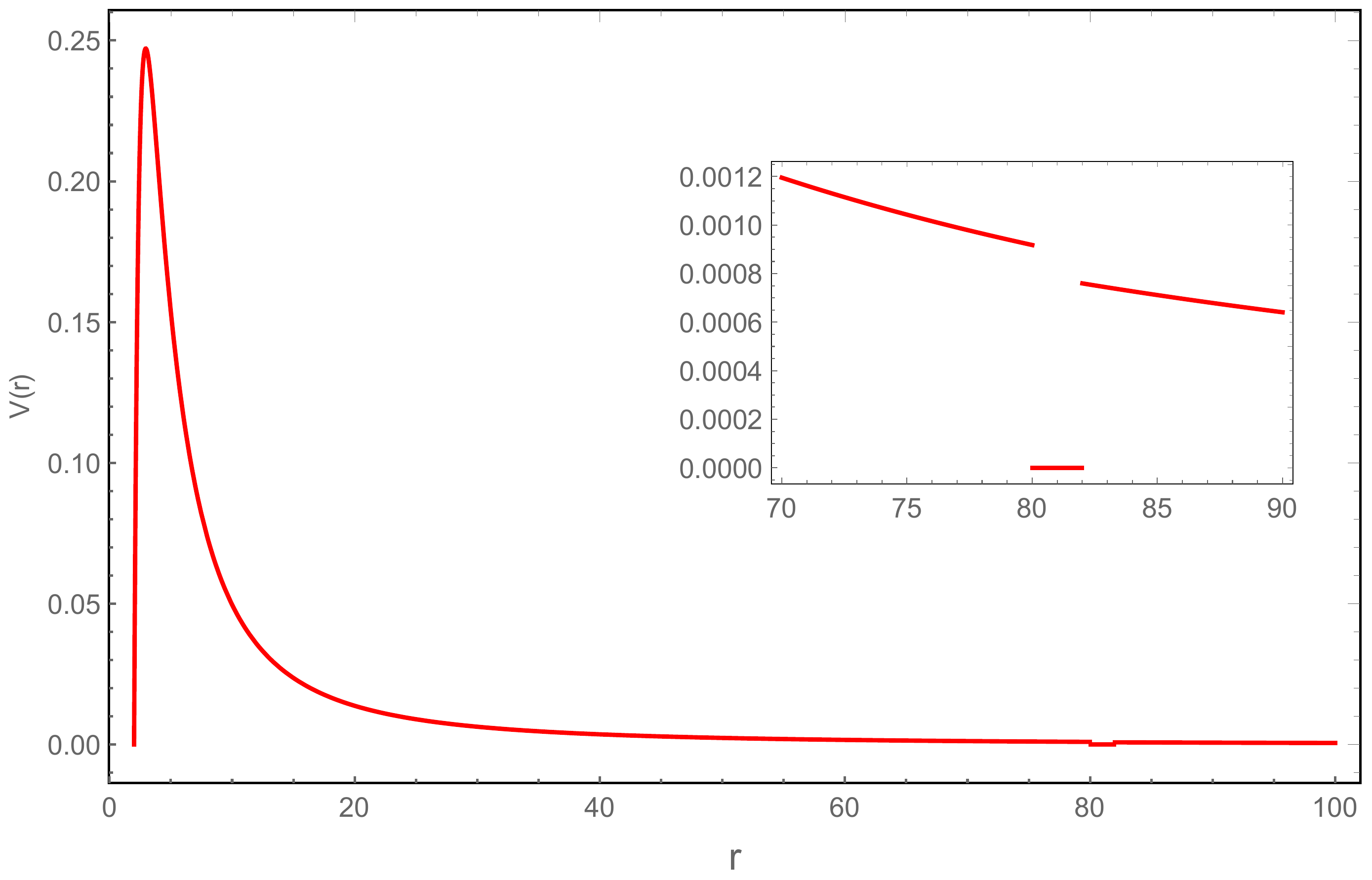}
\includegraphics[height=2in,width=3in]{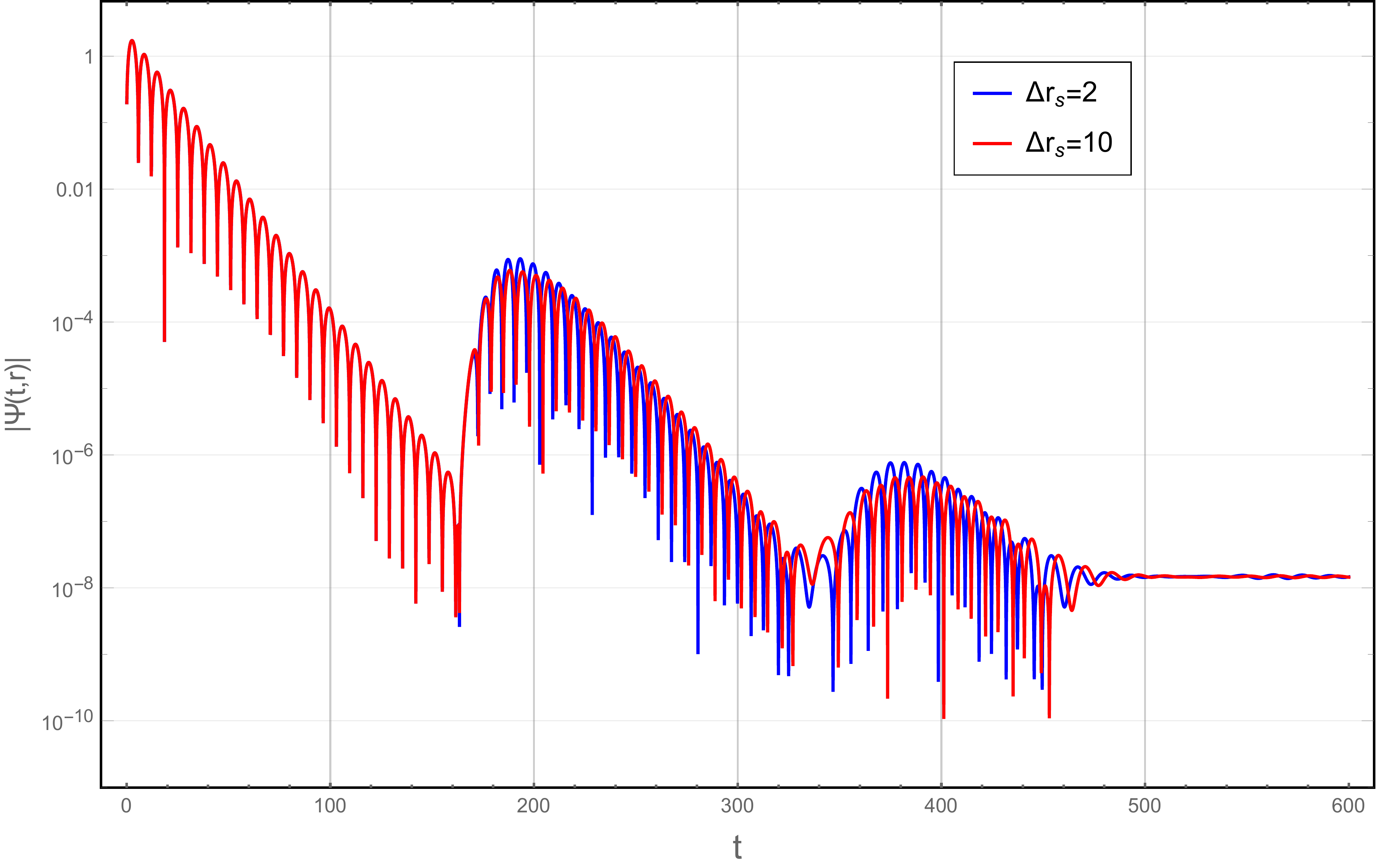}
\caption{
Top: The effetive potential given in Eq.~\eqref{eq2V}, where we take $r_s=80$, $\Delta r_s=2$, $l=2$, and $M_0=1$.
Bottom: The corresponding time-domain profile of scalar field perturbations, where a second curve in purple corresponds to $\Delta r_s=10$..}\label{fig4}
\end{figure}

Intuitively, when compared to the matter distribution defined by Eq.~\eqref{eqm1}, a more straightforward and physically simple option is to replace the distribution with an infinitely thin mass shell.
We may rather adopt a slightly more general scenario where the mass $\Delta M$ is distributed in the range $r_s<r\leq r_s+\Delta r_s$, in which the effective potential vanishes.
This indicats the following effective potential containing a downward ``step'' at $r=r_s$,
\begin{equation}\label{eq2V}
V_2(r) =
\begin{cases}
 \left(1-\frac{2M_0}{r}\right)\left(\frac{l(l+1)}{r^2}+\frac{2M_0}{r^3}\right)   &  r\leq r_s, \\
 0 &  r_s<r\leq r_s+\Delta r_s,\\
 \left(1-\frac{2(M_0+\Delta M)}{r}\right)\left(\frac{l(l+1)}{r^2}+\frac{2(M_0+\Delta M)}{r^3}\right),  & r_s+\Delta r_s<r,
\end{cases}
\end{equation}
which corresponds to the case that the black hole is surrounded by a mass shell $\Delta M$ with finite thickness $\Delta r_s$.
It is not difficult to derive the corresponding mass function $m_2(r)$, which reads
\begin{equation}\label{eq8}
m_2(r) =
\begin{cases}
M_0,  &  r\leq r_s, \\
\frac{1}{2}\left[l(l+1)\ln\left(\frac{r}{r_s}\right)+\frac{2M_0}{r_s}\right]r, &  r_s<r\leq r_s+\Delta r_s,\\
M_0+\Delta M,  &  r_s+\Delta r_s<r,
\end{cases}
\end{equation}
where
\begin{equation}
\Delta M=\frac{1}{2}\left[l(l+1)\ln\left(\frac{r_s+\Delta r_s}{r_s}\right)+\frac{2M_0}{r_s}\right](r_s+\Delta r_s)-M_0 .\label{deltaMass}
\end{equation}

The resultant effective potential and the time-domain profiles are shown in Fig.~\ref{fig4}.
It is observed that the temporal evolutions of the scalar field, regarding two different effective potentials of different thickness, turn out to be largely the same.
According to Eq.~\eqref{deltaMass}, the size of the ``step'' is associated with its width, and therefore the above results indicate that the echoes are not sensitive to the size of the discontinuity.

\subsection{Echoes in potential function with $\mathcal{C}^1$ discontinuity}

In the previous subsection, we have demonstrated that the echoes can be produced in potentials with a discontinuity planted away from the horizon.
The planted discontinuity can be either a ``cut'' or a ``step'', and the results are robust in the sense that they are not sensitive to the strength or location of the discontinuity.
In both cases, however, the potential functions in question are not continuous ones.

It was shown~\cite{Qian:2020cnz} that a discontinuity will significant modify the asymptotical quasinormal mode spectrum.
Moreover, the occurrence of such modification does not depend on the order of the discontinuity.
Therefore, it might be interesting to verify whether it is also the case regarding the black hole echoes.

Accordingly, we consider a mass distribution which corresponds to a $\mathcal{C}^{0}$ continuous but $\mathcal{C}^{1}$ discontinuous effective potential.
A simple choice, which has been introduced in~\cite{Barausse:2014tra}, reads
\begin{equation}\label{eq9}
m_3(r) =
\begin{cases}
   M_0,  &  r < r_s, \\
   M_0+\Delta M\left(3-2\frac{r-r_s}{\Delta r_s}\right)\left(\frac{r-r_s}{\Delta r_s}\right)^2, &  r_s\leq r\leq r_s+\Delta r_s,\\
   M_0+\Delta M,  &  r_s+\Delta r_s<r.
\end{cases}
\end{equation}
It is readily seen to give the following effectively potential, which adquately serves our purpose,

\begin{equation}\label{eq3V}
V_3(r) =
\begin{cases}
 \left(1-\frac{2M_0}{r}\right)\left(\frac{l(l+1)}{r^2}+\frac{2M_0}{r^3}\right)   &  r< r_s, \\
 V_{\mathrm{sh}}  &  r_s\leq r\leq r_s+\Delta r_s, \\
 \left(1-\frac{2(M_0+\Delta M)}{r}\right)\left(\frac{l(l+1)}{r^2}+\frac{2(M_0+\Delta M)}{r^3}\right),  & r_s+\Delta r_s<r,
\end{cases}
\end{equation}
where
\begin{equation}
\begin{split}
V_{\mathrm{sh}}&=\frac{1}{r^4\Delta r_s^6}\left[(4(r-r_s)^3\Delta M-6(r-r_s)^2\Delta M \Delta r_s\right.\\
&\left. +(r-2M_0)\Delta r_s^3)(4(r-r_s)^2(2r+r_s)\Delta M\right.\\
&\left. -6(r^2-r_s^2)\Delta M\Delta r_s+(2M_0+l(l+1)r)\Delta r_s^3)\right] .
\end{split}
\end{equation}

\begin{figure*}
\centering
\includegraphics[height=2in,width=3in]{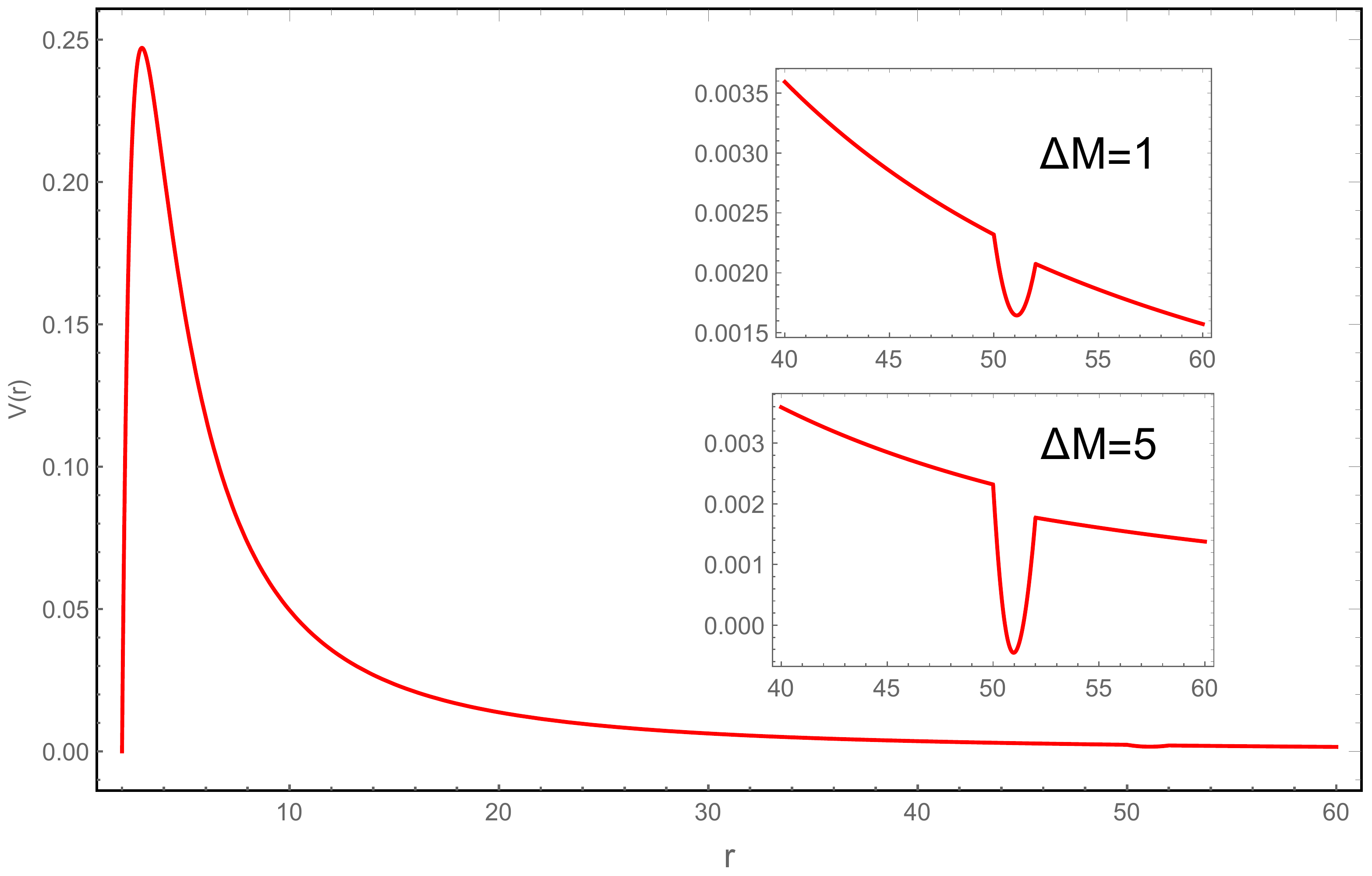}
\includegraphics[height=2in,width=3in]{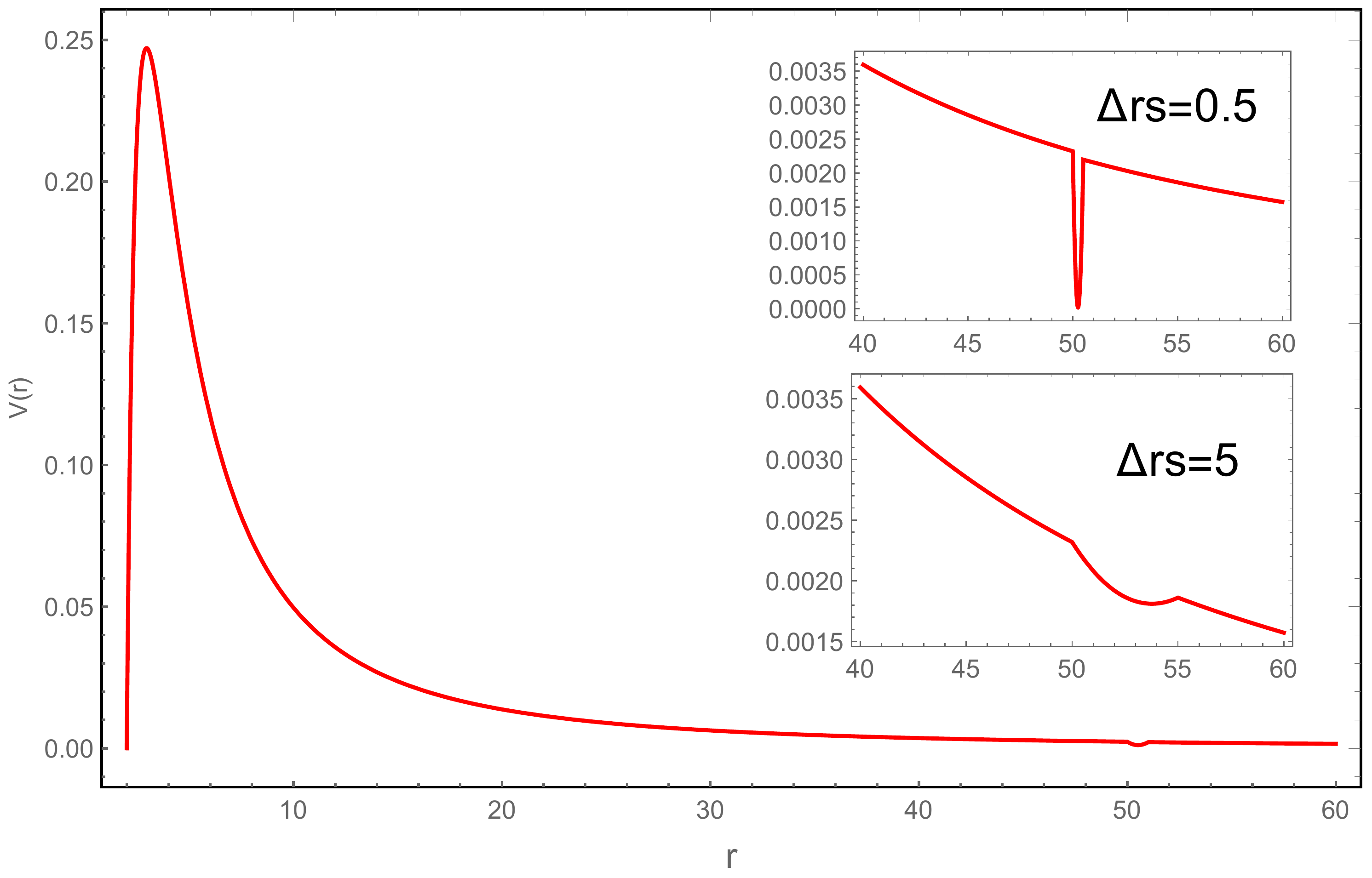}
\caption{The effetive potential given in Eq.~\eqref{eq3V}, where we take $r_s=50$, $l=2$, $M_0=1$.
In the left panel, we choose $\Delta r_s=2$ with $\Delta M=1$ and $\Delta M=5$, while in the right panel, we choose $\Delta M=1$ with $\Delta r_s=0.5$ and $\Delta r_s=5$.}\label{fig5}
\end{figure*}

In Fig.~\ref{fig5} we show the effective potential Eq.~\eqref{eq3V} by varying its parameters.
It is observed, as long as $r_s\gg r_{\mathrm{ph}}$, the change in the global feature of the effective potential is hardly noticeable, even though the parameters have varied significantly.

\begin{figure*}
\centering
\includegraphics[height=2in,width=3in]{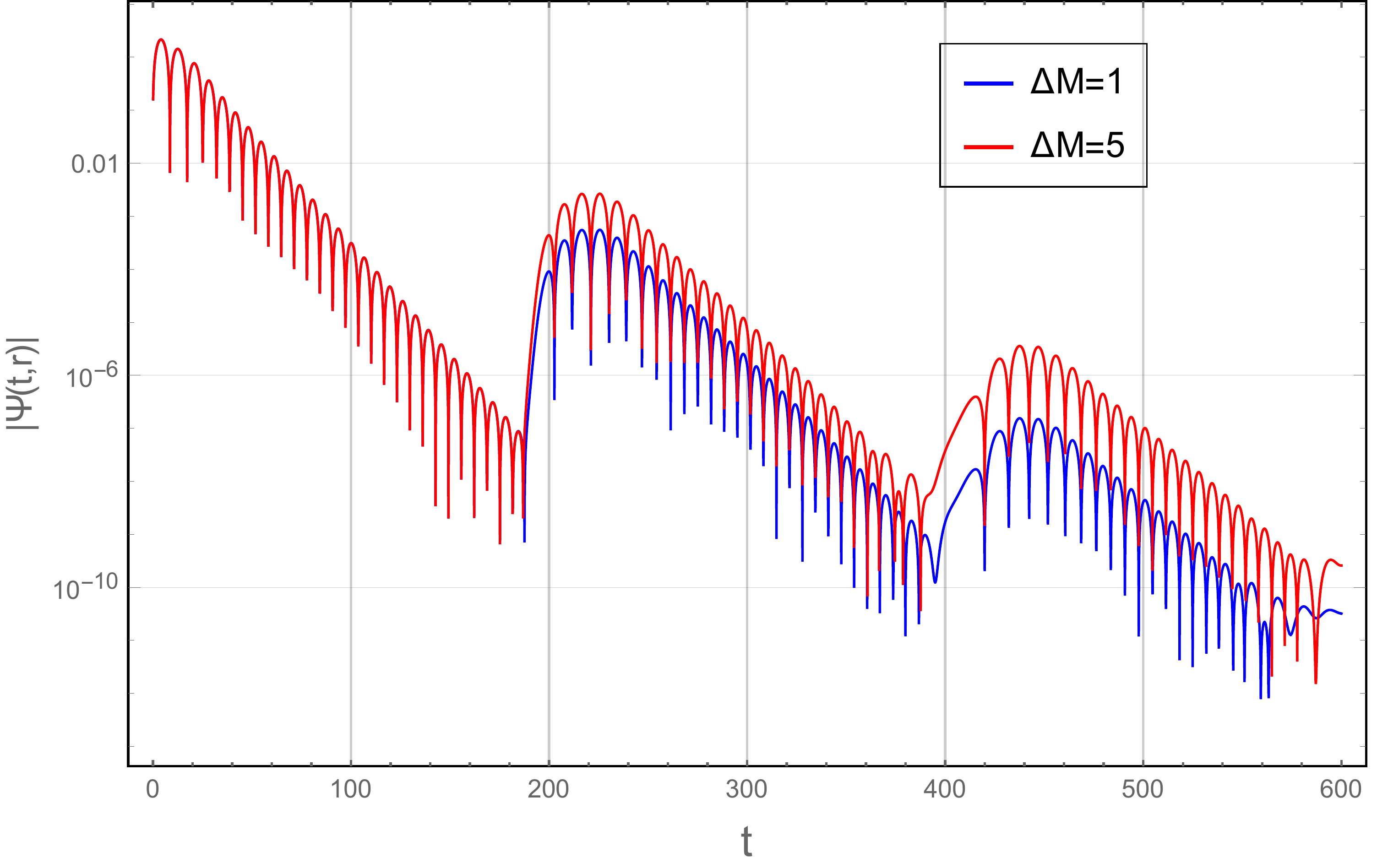}
\includegraphics[height=2in,width=3in]{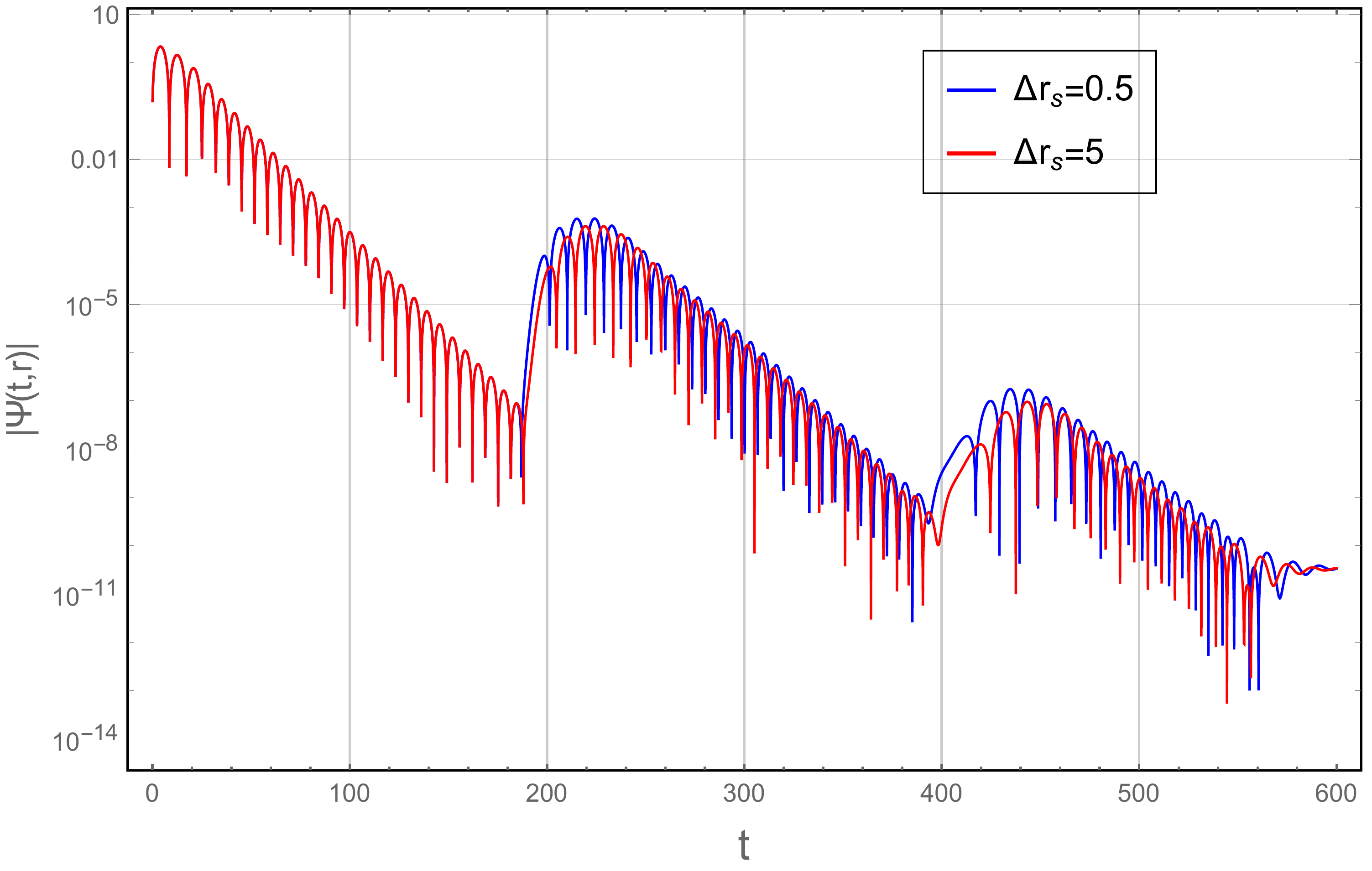}
\caption{The time-domain profile of scalar field perturbations for the effective potential $V_3$ defined by Eq.~\eqref{eq3V}.
The calculations are carried out by taking $l=2$, $M_0=1$, and $r_s=100$.
In the left panel, we show the results by choosing $\Delta r_s=2$ with $\Delta M=1$ and $\Delta M=5$, while in the right panel, we choose $\Delta M=1$ with $\Delta r_s=0.5$ and $\Delta r_s=5$.}\label{fig6}
\end{figure*}

The corresponding time-domain profiles of the scalar perturbations are presented in Fig.~\ref{fig6}.
As expected, the echoes are again observed.
These results further demonstrate the robustness of the proposed picture.
Even though the size and strength of the discontinuity vary significantly in terms of $\Delta r_s$ and $\Delta M$, the echoes persist and seem rather insensitive to those parameters.

\subsection{A comparison between different echo mechanisms}

In the previous subsections, we have shown that the discontinuity in the effective potential may also serve as an alternative mechanism for black hole echoes.
Therefore, it seems rather intriguing to ask whether one could distinguish between the echo patterns associated with different origins from an observational viewpoint.
This would be meaningful, especially if echoes can be established in data.
To this end, in this subsection, we carry out the following numerical study aiming at a comparison between different echo mechanisms.
In particular, departing from the original Schwarzschild black hole with mass $M_0$, we consider two scenarios where the corresponding effective potentials are mostly identical.
On the one hand, one introduces a discontinuity as that given in Eq.~\eqref{eq3V}.
On the other, one devises a wormhole metric by the {\it cut and paste} procedure applied to the Regge-Wheeler potential at the throat $r_0>2M_0$, following~\cite{Cardoso:2016rao}.
We consider two different sets of parameters.
For the first one, we intentionally choose the parameters so that the relevant scales of the physical systems are largely identical.
To be specific, the period of the echoes for the wormhole is governed by the separation between the two maxima of the potential, while in the present model the period is determined by the distance between the maximum and the discontinuity.
By tuning the wormhole's throat length and the position of the potential's discontinuity, one matches the resultant periods of echoes.
The motivation to employ such a configuration is that we are then capable of focusing on the difference in echo waveforms. 
Moreover, we also illustrate a second case where, for some reason, the length scales of the underlying physical systems are different.
One may argue that it might be plausible, as pointed out in~\cite{Cardoso:2019rvt}, that the wormhole might not be arbitrarily compact due to considerations regarding the stability and formation process.
Here, we assume that the relevant scale in the wormhole is smaller compared to that related to the discontinuity, for instance, the size of a halo surrounding the black hole in question.

\begin{figure*}
\centering
\includegraphics[height=2.in,width=3.in]{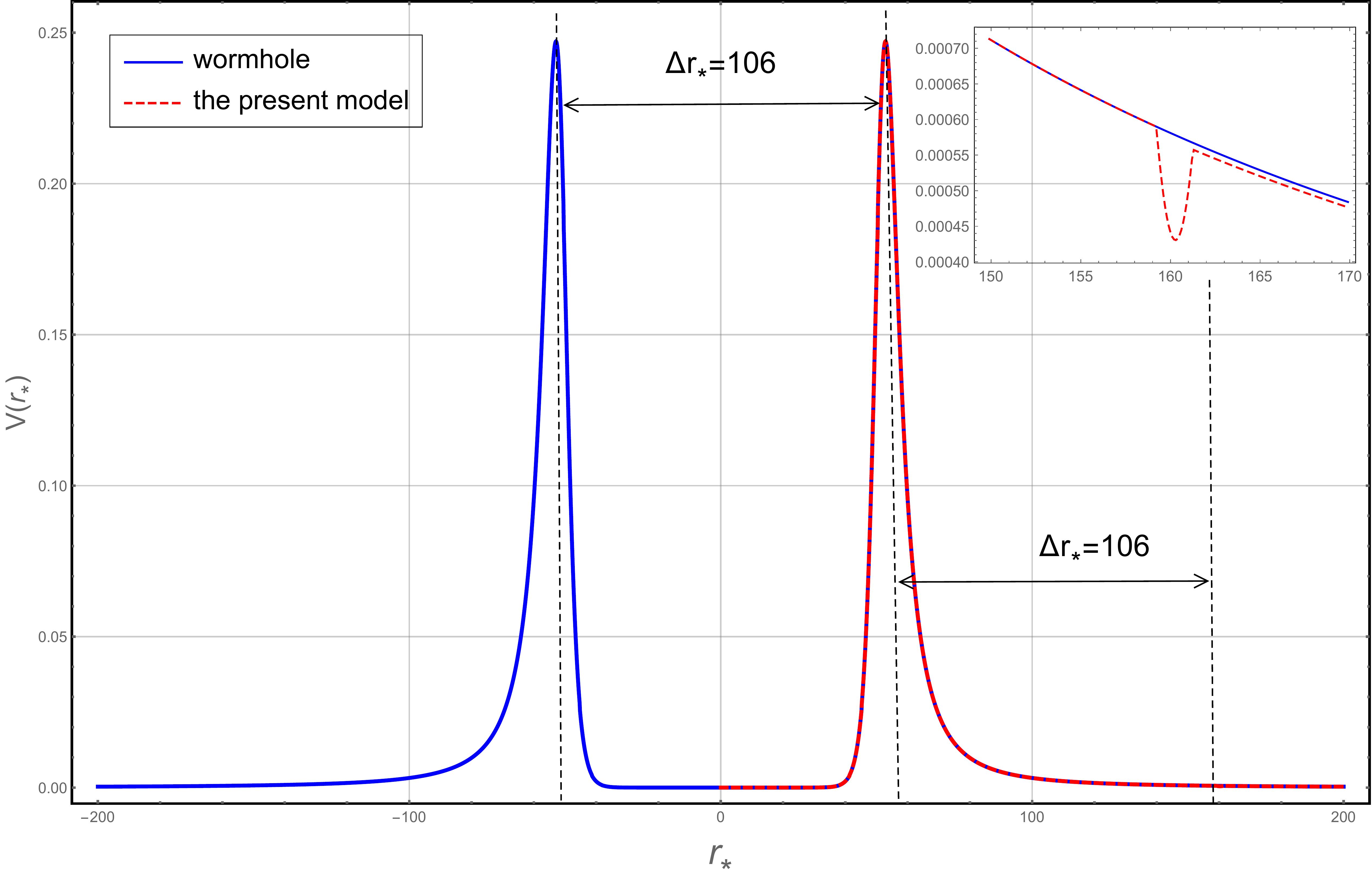}
\includegraphics[height=2.in,width=3.in]{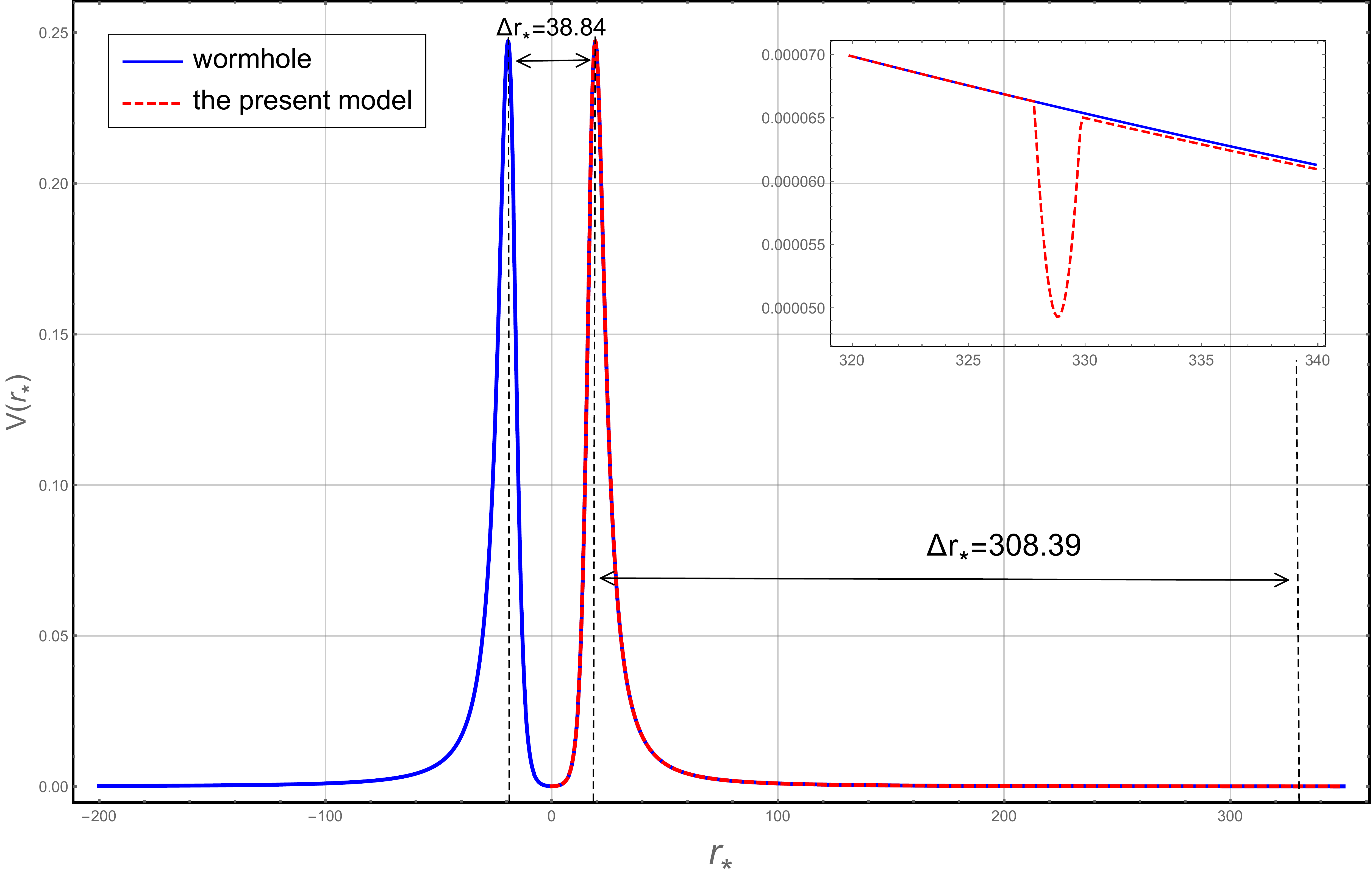}
\caption{The effective potential of the devised wormhole and that given by Eq.~\eqref{eq3V}, shown in the tortoise coordinate.
In the left panel, for the effective potential with discontinuity, we choose $M_0=1$, $l=2$, $r_s=100$, $\Delta r_s=2$, and $\Delta M=1$, and
for the wormhole metric, one takes $r_0=2+0.5\times 10^{-11}$.
Such a choice of parameters guarantees that periods of the echoes are identical in the two cases. 
In the right panel, we take $r_s=300$ and $r_0=2.0001$, while the other parameters remain unchanged.
The black double-sided arrows indicate the relevant scales associated with the bouncing of the waveform.}
\label{potentialCMP}
\end{figure*}

\begin{figure*}
\centering
\includegraphics[height=2.in,width=3.in]{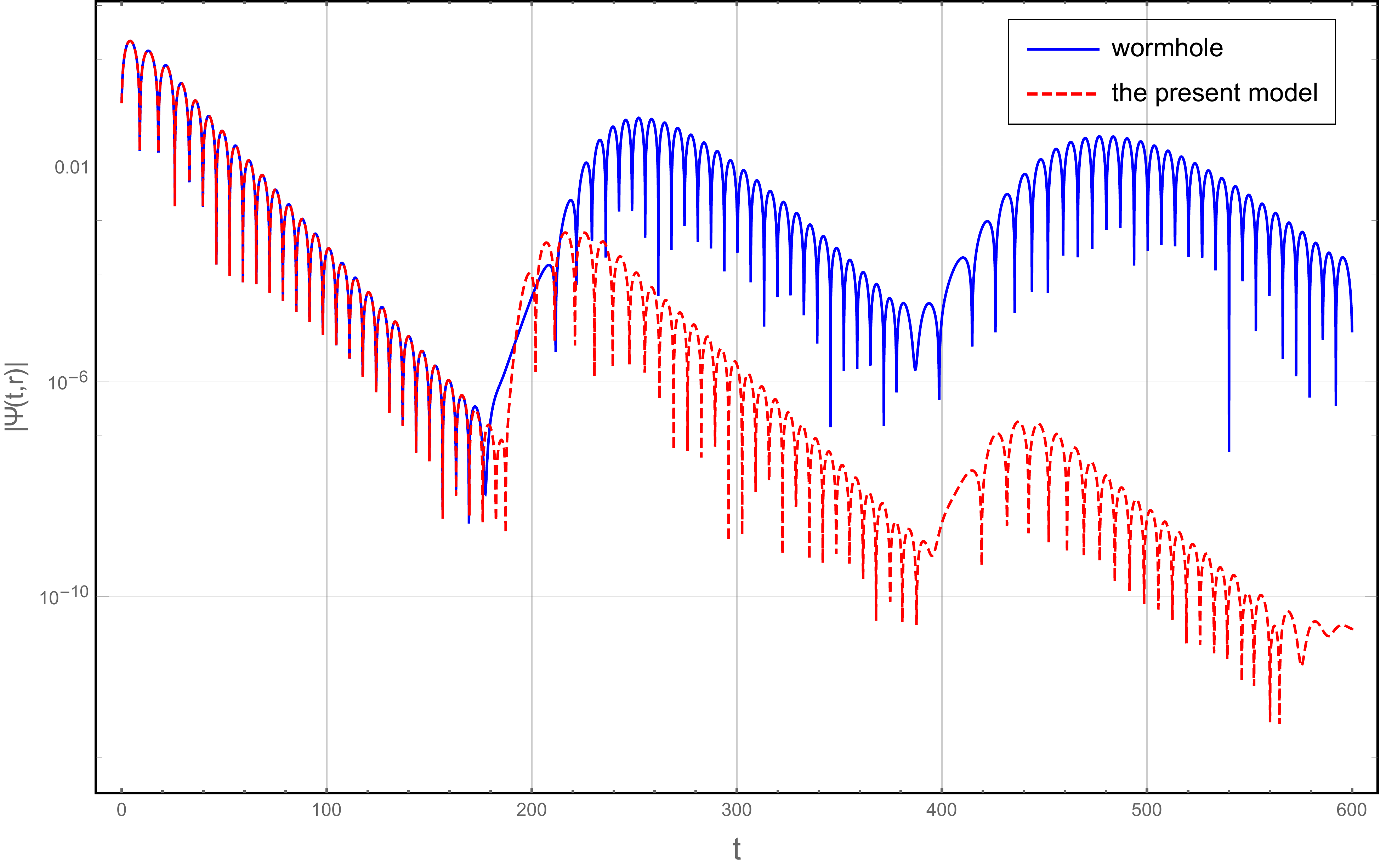}
\includegraphics[height=2.in,width=3.in]{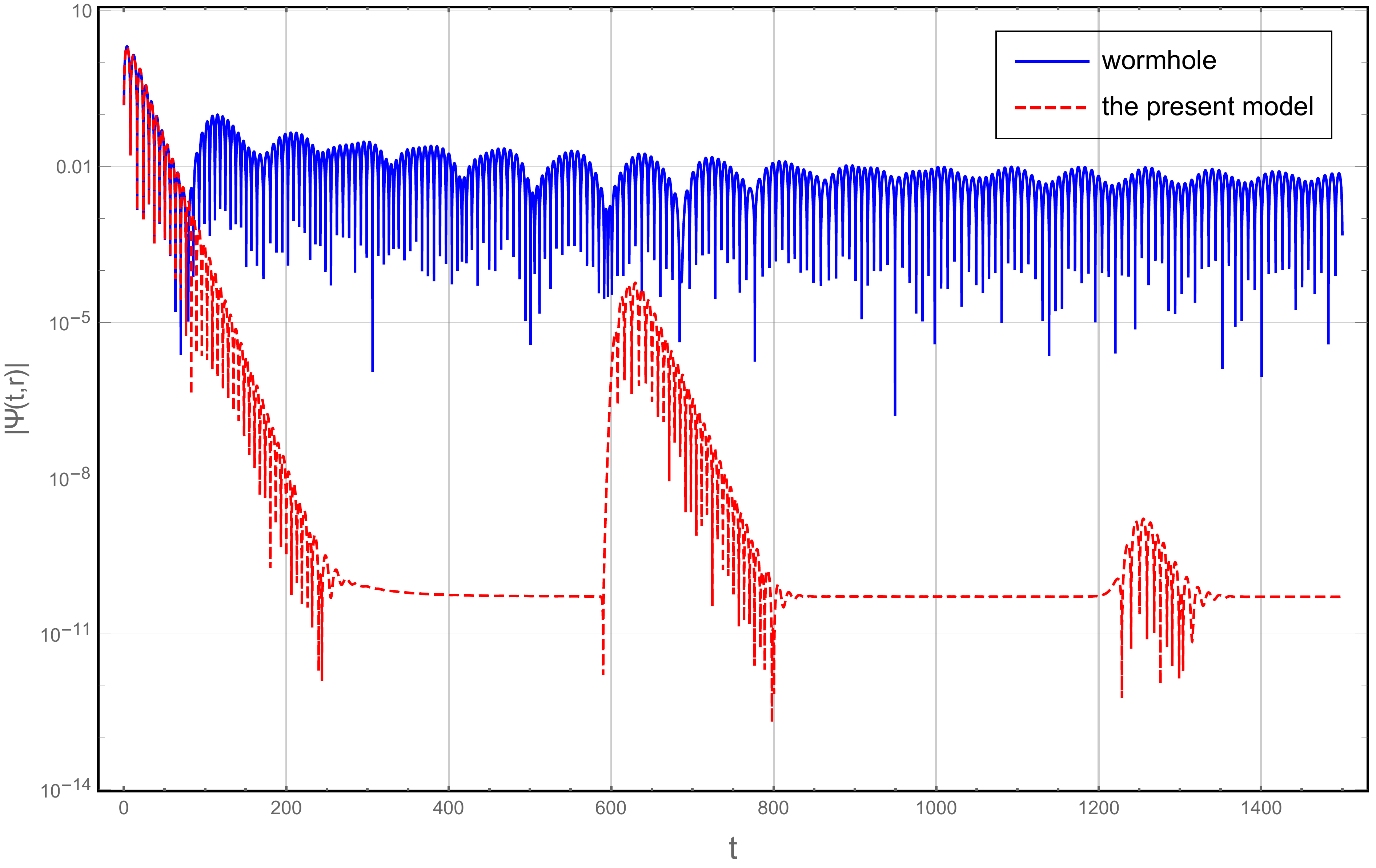}
\caption{Comparisons between the resultant echoes in the time-domain profiles for the two sets of effective potentials shown in Fig.~\ref{potentialCMP}.}
\label{echoesCMP}
\end{figure*}

The corresponding effective potentials are depicted in Fig.~\ref{potentialCMP}.
As shown in both panels of Fig.~\ref{potentialCMP}, for the most part, the resultant effective potentials mostly coincide, and the discontinuity is barely visible.
By feeding the same scalar initial perturbations to the two spacetime configurations, the comparisons between the resultant echoes in the time-domain evolution are given in Fig.~\ref{echoesCMP}. 
One observes two major differences in these scenarios.
First, the maxima of successive echo pulses decrease more rapidly in the case of discontinuity (shown in red dashed curves).
It is clearly indicated by the left panel of Fig.~\ref{echoesCMP}, especially when the echo period are tuned to be identical.
Also, the front of the echo waveform seems distinct between the two cases, while the slope and the main feature of the ringdown part of each echo pulse are mostly indistinguishable.
To be specific, while the echoes in the wormhole are more reminiscent of a wave-packet bouncing back and forth, those due to discontinuity are featured by a sudden rise followed by subsequential ringdown.
Such a difference might be more apparent by observing the comparisons given in both panels of Fig.~\ref{echoesCMP}.
As will be discussed in the next section around Eq.~\eqref{t1new}, the resultant time-domain profile of the present model can be understood as the quasinormal ringdown modulated by a deformed periodic function.
While the main feature of each echo pulse is primarily determined by a few low overtone quasinormal modes of the original black hole, it is subjected to the deformation governed by the Fourier transform of $C_1C_2^2$ defined below by Eqs.~\eqref{Cn}.
The observed difference in the front of the waveform seems to reflect such subtlety.
Besides, from a practical perspective, the strength of the echoes owing to the presence of a discontinuity is less pronouncing and thus would not be straightforward to be detected.
Secondly, if different mechanisms imply different length scales, then it is straightforward that they will be shown in terms of the echo period.
As shown in the right panel of Fig.~\ref{echoesCMP}, the resultant echo periods are rather different, which might provide a distinct signature in the measurement.
We note that, for both cases, the periods of the echoes is consistent with the scales in the effective potentials associated with the bouncing of the waveforms, as indicated by the double-sided arrows.

\section{The connection between the echos and the modified asymptotic quasinormal modes} \label{section4}

In this section, we investigate the echoes from an analytic perspective and explore their relationship with the asymptotical spectrum of the quasinormal modes.
We start by deriving the Green function of the eigenvalue problem defined in Eq.~\eqref{eq2} with appropriate boundary conditions.
With the proper form of the Green function, we proceed further by establishing a relationship with the echoes in the scalar field observed numerically in the previous section.
For simplicity, we focus on the scenario Eq.~\eqref{eq1} where the effective potential is featured by a ``cut''.
It is rather intuitive to generalize our results further to other more sophisticated cases.

According to the standard procedure~\cite{agr-qnm-review-02, agr-qnm-lq-01}, the Green function of the original black hole with the Regge-Wheeler potential reads
\begin{equation}
G(\omega, x,x')= \frac{1}{W(\omega)}f_-(\omega, x_<)f_+(\omega, x_>) ,\label{FormalGreen}
\end{equation}
where $x_<\equiv \min(x, x')$, $x_>\equiv \max(x, x')$, and
\begin{equation}
W(\omega) \equiv W(f_+, f_-) = {f_+}' {f_-} - {f_-}' {f_+} \nonumber
\end{equation}
is the Wronskian of $f_-$ and $f_+$.
Here $f_-$ and $f_+$ are the two linearly independent solutions of the corresponding homogeneous equation satisfying the boundary conditions Eq.~\eqref{master_bc0} at the horizon and infinity.
To be specific, $f_-$ and $f_+$ possess the following asymptotic forms
\begin{equation}
f_-(\omega, x) \sim
\begin{cases}
   e^{-i\omega x}, &  x \to -\infty, \\
   A_{\mathrm{out}}(\omega)e^{+i\omega x}+A_{\mathrm{in}}(\omega)e^{-i\omega x}, &  x \to +\infty,
\end{cases}
\label{master_bc1}
\end{equation}
and
\begin{equation}
f_+(\omega, x) \sim    e^{+i\omega x},\ \ \ \  x \to +\infty ,
\label{master_bc2}
\end{equation}
in asymptotically flat spacetimes, which are bounded at $t\to +\infty$ for $\Im \omega <0$.
In the above expressions, $A_{\mathrm{in}}$ and $A_{\mathrm{out}}$ are the reflection and transmission amplitudes, whose specific forms are unknown to us but are well-defined for a given metric.
It is noted that the asymptotical form of $f_+(\omega, x)$ at $x\to -\infty$ is not given above, as it does not concern our present derivations.

The effective potential of the present problem Eq.~\eqref{eq1} contains a cut, which introduces a discontinuity at the radial coordinate $r=r_s$.
In this case, the formal solution for the Green function Eq.~\eqref{FormalGreen} is no longer valid in the entire range of the tortoise coordinate $-\infty <x <+\infty$.
However, it is still meaningful if we only focus on the region $-\infty <x < x_s(r_s)$.
Accordingly, one of the two solutions for the homogeneous equations $f_+$, as well as the corresponding asymptotic form Eq.~\eqref{master_bc2}, must be modified to adapt to the change.

One observes that physically relevant boundary condition for the modified effective potential is still given at $x\to +\infty$.
As the potential vanishes identically in this region, it is simply the outgoing plane wave
\begin{equation}
g(\omega, x)=e^{+i\omega x} .\label{gBound}
\end{equation}
Now, since the relevant range of the modified solution $\tilde{f}_+$ ends at $x=x_s$, its boundary condition should also be defined at $x=x_s$.
Moreover, in principle, it should be connected to $g(\omega, x)$ in accordance with the master equation.
However, the difficulty is that we do not have a simple analytical form since the exact form of $f_+$ is unknown to us.
In order to proceed further analytically, let us assume that $x_s$ is numerically large, and the potential is rather smooth at $x\lesssim x_s$.
Under the above assumption, as shown in the Appendix, the modified solution of homogeneous equation is found to be
\begin{equation}
\tilde{f}_+(\omega, x) = f_+(\omega, x) + \frac{\mathcal{R}}{A_{\mathrm{in}} e^{-2i\omega x_s}-\mathcal{R}A_{\mathrm{out}}} f_-(\omega, x) , \label{tildeF}
\end{equation}
where $\mathcal{R}$ is given by Eq.~\eqref{defR}.
It is worth noting that a fraction of $f_-(\omega, x)$ have now sneaked into the homogeneous solution $\tilde{f}_+(\omega, x)$.
Although being insignificant as $x\to x_s$, its coefficient will give rise to additional attribution to the Green function, which turns out to be essential to the presence of echoes, as will become apparent shortly.

One can readily evaluate the Green function Eq.~\eqref{FormalGreen} by replacing the solution associated with the original black hole $f_+\to \tilde{f}_+$.
This subsequently modifies the original Green function to give
\begin{equation}
\begin{split}
&G(\omega, x,x')\to \tilde{G}(\omega, x,x')  \\
&=\frac{1}{W(\tilde{f}_+, f_-)}f_-(\omega, x_<)\tilde{f}_+(\omega, x_>) \\
&=G(\omega, x,x')+ \frac{\mathcal{R}}{A_{\mathrm{in}} e^{-2i\omega x_s}-\mathcal{R}A_{\mathrm{out}}}\frac{1}{W(\omega)}f_-(\omega, x_<)f_-(\omega, x_>) \\
&=G(\omega, x,x')+ \frac{\mathcal{R}}{A_{\mathrm{in}} e^{-2i\omega x_s}-\mathcal{R}A_{\mathrm{out}}}\frac{1}{W(\omega)}f_-(\omega, x)f_-(\omega, x') , \label{FormalGreenCut}
\end{split}
\end{equation}
where $G(\omega, x,x')$ is the Green function of the original black hole metric.

\begin{figure}
\centering
\includegraphics[height=2in,width=3in]{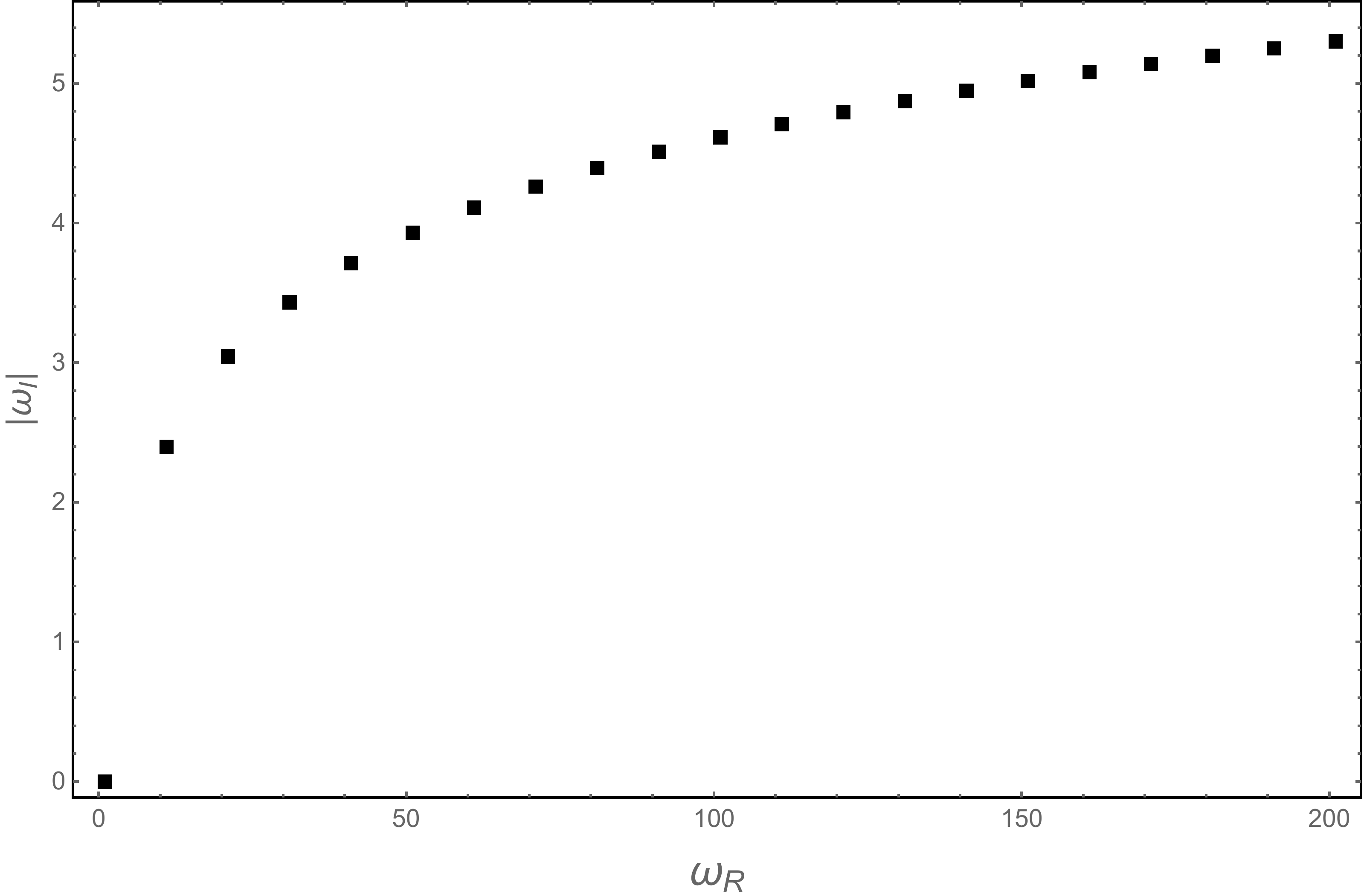}
\caption{A schematic plot of the asymptotical behavior of the quasinormal modes frequencies whose distribution in the complex plane are approximately along the real
axis.}\label{fig7}
\end{figure}

It is noted that the Wronskian $W(\omega)$ in Eq.~\eqref{FormalGreen} remains unchanged since the second term on the r.h.s. of Eq.~\eqref{tildeF} does not bring any contribution.
As a result, the pole structure associated with the original black hole metric will still be present in both terms on the r.h.s. of Eq.~\eqref{FormalGreenCut}.
The novelty comes from the denominator of the additional factor, namely, ${A_{\mathrm{in}} e^{-2i\omega x_s}-\mathcal{R}A_{\mathrm{out}}}$.
It is not difficult to show that it corresponds to a spectrum of poles that lies approximately along the real axis of the frequency~\cite{Qian:2020cnz}.
This can be understood as follows.
For simplicity, let us first assume that the amplitudes $A_{\mathrm{in}}$, $A_{\mathrm{out}}$, and $\mathcal{R}$ are constants.
Subsequently, if one finds that the factor vanishes for a certain frequency $\omega$, then adding a multiple of $\pi/x_s$ to the real part of $\omega$ will yield another root.
In practice, these amplitudes (or the relevant ratios between them) are not constants but moderate functions of the frequency $\omega$, especially when compared to the exponential function $e^{-2i\omega x_s}$.
As a result, the above justification still largely holds valid at the limit where the frequencies possess large real part $\Re\omega \gg 1$ and modest negative the imaginary part $\Im\omega <0$.
The asymptotic spectrum of frequencies lines up alone the real axis with an interval of $\Delta(\Re\omega)=\pi/x_s$ between successive poles, while their imaginary parts do not vary significantly as illustrated in Fig.~\ref{fig7}.

Now, we proceed to demonstrate that the above asymptotic properties of the quasinormal modes are closely related to the echoes observed in the last section.
This can be accomplished by studying the time-domain evolution of some initial perturbations.
If the latter is moderate in the sense that it does not contain any pole~\cite{agr-qnm-lq-01}, the temporal evolution is determined by the inverse Fourier transform of Eq.~\eqref{FormalGreenCut}.
The resonant frequencies are obtained when one encloses the counter along a large circle in the lower half of the complex plane, and subsequently, the result is essentially governed by the poles of the Green function.
It is found that the first term of the r.h.s. of Eq.~\eqref{FormalGreenCut} leads to identical quasinormal oscillations of the original black hole metric.
The second term, on the other hand, is a product of two factors, namely,
\begin{equation}
\frac{\mathcal{R}}{A_{\mathrm{in}} e^{-2i\omega x_s}-\mathcal{R}A_{\mathrm{out}}} \nonumber
\end{equation}
and
\begin{equation}
\frac{1}{W(\omega)}f_-(\omega, x)f_-(\omega, x') . \nonumber
\end{equation}
The resultant temporal evolution is thus governed by the convolution of the two functions resulting from the corresponding inverse Fourier transforms.

In order to show the content of the first factor more transparently, let us again, for the moment, assume that $A_{\mathrm{in}}$, $A_{\mathrm{out}}$, and $\mathcal{R}$ are constants.
we note that the inverse Fourier transform of the first term gives
\begin{equation}
F(t) = C_1\delta(t) + C_1 C_2 \delta(t+T) + C_1 C_2^2 \delta(t+2T) + C_1 C_2^3 \delta(t+3T) + \cdots  , \label{deltaSumFormal}
\end{equation}
where
\begin{equation}
\begin{split}
C_1 &= -\frac{\sqrt{2\pi}}{A_{\mathrm{out}}}  \\
C_2 &=\frac{A_{\mathrm{in}}}{A_{\mathrm{out}}\mathcal{R}} , \\
T &= 2x_s . \label{Cn}
\end{split}
\end{equation}
The physical interpretation of the second factor is mostly determined by its denominator.
As the Wronskian stays the same, it possesses the same quasinormal frequencies of the black hole metric.
%In particular, if the original effective potential is the P\"oschl-Teller one, namely, symmetric about its maximum, the second factor will be identical to the original Green function.
Therefore, its inverse Fourier transform, $H(t)$, is characterized by dissipative oscillations which, to a large extent, are reminiscent of the original black hole quasinormal modes.

By putting all the pieces together, one is ready to evaluate the convolution of Eq.~\eqref{deltaSumFormal} with $H(t)$ and finds
\begin{equation}
\begin{split}
&\tilde{G}(t)\sim \int d\tau F(\tau)H(t-\tau) \\
&= C_1H(t) + C_1 C_2 H(t+T) + C_1 C_2^2 H(t+2T)  + \cdots  , \label{deltaConvFormal}
\end{split}
\end{equation}
where the resultant time evolution is featured by superposition of individual pulses.
Moreover, the pulses are modulated by identical templates proportional to the quasinormal oscillations $H(t)$, with the interval $T$ between successive ones.
One observes that the period $T$ is nothing but the time interval for the wave to travel to and be bounced back from the ``cut'' at $x_s$, consistent with our earlier numerical results.
From the above discussions, it is apparent that $T$, the spacing in the time-domain, is intrisically related to the distance between successive poles along the real axis $\Delta(\Re\omega)$ in the frequency domain, by the following relation
\begin{equation}
\Delta(\Re\omega) = 2\pi/T . \label{Arelation}
\end{equation}
Therefore, the echoes observed in Eq.~\eqref{deltaConvFormal} can be traced back to precisely the same origin, from which the asymptotic spectrum of quasinormal modes~\cite{Qian:2020cnz} also arises.
Further discussions regarding the relationship between the findings in~\cite{Qian:2020cnz} and the present study will be given in the last section.

Now, let us roll back and reiterate the above derivations in more rigorous terms.
One may rewrite the first factor as
\begin{equation}
\frac{\mathcal{R}}{A_{\mathrm{in}} e^{-2i\omega x_s}-\mathcal{R}A_{\mathrm{out}}} = \frac{C_1}{\sqrt{2\pi}} \sum\limits_{j=0} \left(C_2 e^{-i T\omega} \right)^j . \label{t1new}
\end{equation}
Since $C_1$ and $C_2$ are functions of the frequency, they cannot be treated as constant in an inverse Fourier transform.
However, for each term in the summation, we can still attach the $C_1C_2^i$ term to the second factor, and carry out the inverse Fourier transform of the result expression.
As the poles governed by the Wronskian will not be modified, even though the pulse will be deformed, the main feature will maintain.
The above results, therefore, largely remain valid.
In particular, regarding the relation Eq.~\eqref{Arelation}, the period of the echoes obtained above stays unchanged.
The spacing between successive poles of the Green function in the frequency domain, on the other hand, should be replaced by its asymptotic value at large real frequency, namely, $\lim\limits_{\Re\omega\to+\infty}\Delta(\Re\omega)$.
We therefore have
\begin{equation}
\lim\limits_{\Re\omega\to+\infty}\Delta(\Re\omega) = 2\pi/T . \label{ASrelation}
\end{equation}
Before closing this section, we note that although the physical setup of the problem is different, somehow, very similar conclusions have been drawn in~\cite{agr-qnm-echoes-15}.
We relegate further discussions regarding this aspect to the next section.

\section{Further discussions and concluding remarks} \label{section5}

In~\cite{agr-qnm-echoes-15}, Mark {\it et al.} explored the black hole echoes through the analysis of the properties of the Green function.
The authors considered a simplified problem that applies to sufficiently compact ECOs and all relevant sources reside in the Schwarzschild portion of the metric.
In this case, the incident wave is reflected at the ECO surface $r_0$ which gives rise to a modified boundary condition when compared to that of the black hole.
The Green function is constructed by adding to the black hole Green function a solution of the corresponding homogeneous equation, where an arbitrary coefficient of the latter is tuned to adapt to the modified boundary condition.
This is apparently different from the scenario investigated in the present work where the ingoing wave is admitted at the black hole horizon.
Although the spacetime is modified further away from the horizon, at $r=r_s$, the boundary condition at infinity, namely, the outgoing wave given by Eq.~\eqref{master_bc0}, remains unchanged.
Therefore, it still furnishes an eigenvalue problem in a non-Hermitian system, in other words, the typical setup for the black hole quasinormal modes.
The Green function is subsequently obtained using the formal solution where the modified Wronskian is evaluated.
Further discussions regarding the difference have been carried out in~\cite{agr-qnm-echoes-14}.
Nonetheless, apart from the distinct nature of the relevant physical scenarios, similar conclusions about the echoes have been drawn in both studies.
The temporal evolution is featured by a series of echoes consisting of dissipative oscillations, which are essentially governed by the quasinormal frequencies of the corresponding black hole metric.
The successive echoes are separated by a constant delay, which is largely associated with the time interval for the wave to travel and bounce back between the two maxima of the effective potential.
A subtlety in our case, however, is that one of the maxima is replaced by an insignificant ``cut''.
Moreover, the resultant Green function Eq.~\eqref{FormalGreenCut} bears a strong resemblance to that obtained in~\cite{agr-qnm-echoes-15}.
In this context, we conclude that the underlying mechanism behind the echoes in two distinct physical setups is, by and large, equivalent.

Also aimed at the connection of quasinormal oscillations and the echoes, more specific spacetime configurations have also been studied.
Using a modified P\"oschl-Teller potential with a reflecting wall, Price and Khanna~\cite{agr-qnm-echoes-14} observed that in a sequence of echoes, later echoes are not copies of the proceeding ones.
This can be readily understood by observing Eq.~\eqref{t1new}.
When inversely transformed into the time-domain, each term in the sequence is subjected to the convolution of the delta function with different temporal functions.
As a result, the oscillation pattern of each pulse might be distinct, while the interval between successive echoes remains the same.

Amid numerical investigation of the temporal evolution of initial perturbations, the effect of mass shell was discussed by Barausse {\it et al.}~\cite{Barausse:2014tra} and Konoplya {\it et al.}~\cite{Konoplya:2018yrp}.
In both studies, the authors observed the first pulse of the echoes by considering a finite layer of mass shell.
%However, the authors did not discuss the mathematical origin either relate the observed echoes to the discontinuity in the metric.
The present study, on the other hand, explore further the role of discontinuity, both numerically and analytically.
Numerically, the recurrence of echoes can be clearly identified in Figs.~\ref{fig1}, \ref{fig4}, and~\ref{fig5}, even though the planted modifications in the effective potential are rather insignificant.
In the time domain, the resultant echoes can be viewed as a particular type of amplitude modulation of the existing quasinormal oscillations of the corresponding black hole metric.
To some extent, such behavior is reminiscent of the beat in acoustics, which is originated from the interference between two slightly different frequencies.
In the present case, in the place of two different frequencies, one deals with two branches of complex frequencies.
We note that the emergence of two quasinormal mode spectra in discontinous effective potential has also been noticed recently by Daghigh {\it et al.} in~\cite{agr-qnm-22}.
In the context of the present study, the first spectrum is related to the original black hole metric, which gives rise to dissipative oscillation waveform, mostly determined by the first few low overtone quasinormal frequencies.
The second one, whose asymptotic properties have been recently investigated in~\cite{Qian:2020cnz}, is owing to the modification planted into the metric.
As discussed in~\cite{Qian:2020cnz}, the subsequential modification to the quasinormal mode spectrum is closely associated with the discontinuity of the metric, while irrelevant to the detailed form of the continuous part of the effective potential.
Regarding the black hole echoes, instead of few dominant modes, it is the entire asymptotical spectrum that plays the role to modulate the echoes.
This is because, as illustrated in Fig.~\ref{fig7}, as the magnitudes of the imaginary parts of successive modes are similar, the contribution of a mode of higher overtone is still significant.
As a result, they contribute collectively, which eventually leads to the resultant amplitude modulation.
Therefore, the echoes that modulate the quasinormal oscillations are attributed to an interplay between the asymptotic behavior of the second spectrum and the first few quasinormal modes of the black hole.
As most of the discussions have been carried out without referring to specific details of the metric, we argue that the present findings are physically meaningful on rather general grounds.

It is worth noting that the present findings, which is largely based on the presence of the discontinuity in the effective potential, is physically meaningful only when it can be put into a realistic context.
In this regard, we further elaborate on a few plausible physical scenarios where discontinuity plays a pertinent role.
Firstly, discontinuity often makes its presence at the surface of a star, which subsequently leads to $w$-mode, a family of quasinormal modes found in pulsating stars~\cite{agr-qnm-star-07}.
The physical significance of the $w$-mode is that it does not possess any Newtonian analog.
This is because the dissipative resonance is primarily owing to the spacetime vibrations weakly coupled to those of the fluid constituting the star.
In the literature, such modes are mostly investigated numerically, and one of the essential features of the model is that the star usually possesses a ``hard'' surface with $C^1$ discontinuity.
This is a natural consequence as one considers a star whose matter distribution is confined inside a spatially finite region.
As pointed out in Ref.~\cite{agr-qnm-lq-02}, the asymptotical quasinormal mode spectrum for such a setup can be derived analytically.
Moreover, as the effective potential of such a spacetime configuration is featured by a maximum and a discontinuity, based on the arguments given in the present study, echoes are expected in the temporal evolution of the metric perturbation.
This is indeed the case, though not explicitly referred to as echoes, dampened repetitions of the primary signal have been observed in numerical simulations~\cite{agr-qnm-star-09,agr-qnm-star-11,agr-qnm-star-13}.
In the present study, we argued that the origin of such echoes can be attributed to the discontinuity, and their recurrence interval is closely related to the asymptotic spectrum of the quasinormal modes.
Secondly, discontinuity constitutes an important assembly component in the context of ECO, such as the gravastar~\cite{agr-eco-gravastar-02, agr-eco-gravastar-03} and wormhole~\cite{agr-wormhole-10}.
As a typical horizonless ECO, the gravastar is characterized by non-perturbative corrections to the near-horizon external geometry of the corresponding black hole metric.
In the original picture proposed by Mazur and Mottola, it is implemented by introducing different layers of matter compositions with distinct equations of state, and therefore, it naturally leads to $C^1$ discontinuity on the interfaces.
The construction is subsequently adopted by most generalizations of the model.
In particular, in the analysis of dynamic stability of the spacetime configuration carried out by Visser and Wiltshire, the thin shell in the original model is further simplified to an infinitesimally thin layer with $C^0$ discontinuity.
Also, the concept of the thin shell is essential to construct the throat of traversable wormholes using the cut-and-paste procedure, which allows one to confine exotic matter in a limited part of the spacetime.
Echoes in such ECOs have been extensively discussed by other authors and, in particular, those in gravastar are viewed to encode important information on the possible existence of such black hole alternatives.
The third example is that discontinuity also makes its appearance from a dynamic perspective.
For instance, in dark halos, $C^1$ discontinuity was observed.
In the framework of $\Lambda$CDM model, discontinuity in the halo profile, dubbed ``cusp'', was observed in the N-body numerical calculations~\cite{agr-dark-matter-06, agr-dark-matter-07}.
Although cuspy dark halos are generally viewed as a ``problem'' of the $\Lambda$CDM model, it was also pointed out that the rotation curves of specific galaxies are largely compatible with the presence of cuspy dark halos~\cite{agr-dark-matter-08}.
In the outer region, on the other hand, a sudden drop in the density profile is also observed~\cite{agr-dark-matter-21, agr-dark-matter-24}, referred to as ``splashback'' in the literature.
Physically, this is a boundary associated with the {\it caustic} formed by matter at their first apocenter after infall.
To be more specific, it separates the matter that has already collapsed and orbiting the halo at least once and those are still infalling onto the halo for the first time.
Subsequently, it gives rise to a discontinuity in the outskirt of the halo profile.
Apparently, if a black hole is surrounded by a dark halo with discontinuity, it is subjected to the type of echoes discussed in the present study.
Also, in the study of the time evolution of a spherical collapsing matter when back reaction regarding evaporation is taken into consideration~\cite{agr-collapse-thin-shell-11,agr-collapse-thin-shell-12,agr-collapse-thin-shell-13}, the interior metric is found to possess discontinuity.
Based on the above discussions, we conclude that discontinuity can be treated as a relevant feature in the astrophysical context.
As pointed out, we note that not all echoes are necessarily related to discontinuity, and the present study is devoted to addressing its possibility as an alternative mechanism.

To summarize, in this work we studied an alternative mechanism for echoes by employing a simplified model with discontinuity.
The spacetime configuration of our approach corresponds to initial scalar perturbations triggered in the vicinity of a black hole surrounded by a thin layer of matter shell.
Subsequently, one investigated the temporal evolution of the perturbations as to how the propagation of the field is affected by the black hole spacetime and the cuspy shell.
Instead of assuming an effective potential with two maxima, we explored the possibility where one of the maxima is replaced by a distant ``cut'', which in turn truncates the power-law tail of Regge-Wheeler potential.
The resultant echoes are investigated both numerically and analytically.
By numerically investigating a variety of different scenarios, it is observed that the echoes are formed.
The latter is shown to be rather independent of the specific configuration, although the strength of the echo signals in our model is weaker than those reported earlier.
By exploring the pole structure of the Green function, we show that the echoes can be understood analytically in terms of the modified asymptotical properties of the quasinormal mode spectrum.
In particular, a relation is established between the period of echoes in the time domain and spacing between successive poles in the frequency domain.
We argue that the present findings are of astrophysical relevance and might introduce some subtlety on the experimental side.
We plan to explore the subject further in future studies.

\section*{Appendix}

In this Appendix, we derive Eq.~\eqref{tildeF} given in the main text.
As discussed before, the physically relevant boundary condition for the modified effective potential should be given at $x\to +\infty$.
Since the potential vanishes identically for $x > x_s$, the proper solution is the outgoing plane wave Eq.~\eqref{gBound}.
On the other hand, the function $f_+(\omega, x)$ is bounded at $x=x_s$ and in principle it can be connected to $g(\omega, x)$ at this point unambiguiously.
In practice, unfortunately, we do not have the exact analytic form of $f_+$.
In order to proceed further analytically, we adopt some moderate assumptions which simplify the calculations but do not modify the essential underlying physics.
In particular, we will consider that $x_s$ is numerically large, and the potential is rather smooth at $x\lesssim x_s$.

Under the above assumptions, let us write down $\tilde{f}_+$ as a linear combination of the two independent solution $f_+$ and $f_-$,
\begin{equation}
\tilde{f}_+(\omega, x) = f_+(\omega, x)+B f_-(\omega, x)
\end{equation}
which is always possible.
It is not difficult to obtain the connection condition at the discontinuity $x=x_s$.
When the WKB approximation is valid on the right-hand side of the ``cut'', the condition is derived in Eq.~(24) of Ref.~\cite{Qian:2020cnz}.
For the present case, one can solve the above equation for the coefficient $D$, while making use of the substitutions
\begin{equation}
\begin{split}
C &= 1, \\
k &= \sqrt{\omega^2-V_s},  \\
x_{\mathrm{cut}} &= x_s,  \\
S_{\mathrm{cut}} &= 0 .
\end{split}
\end{equation}
We subsequently find
\begin{equation}
\tilde{f}_+(\omega, x)\sim e^{+i\omega (x-x_s)}+\mathcal{R}e^{-i\omega (x-x_s)}, \ \ \  x \to x_s ,
\label{master_bc3}
\end{equation}
where
\begin{equation}
\mathcal{R}=\frac{\sqrt{\omega^2-V_s}-\omega}{\sqrt{\omega^2-V_s}+\omega} \label{defR}
\end{equation}
and $V_s=V(x_s)$.
As expected, $\mathcal{R}$ vanishes with $V_s$ when $x_s \to +\infty$.
It is noted that in the special case of constant potential, the reflection amplitude $\mathcal{R}$ can be derived straightforwardly by solving a one-dimensional scattering problem.

By using the specific form of Eq.~\eqref{master_bc3}, one can now identify the coefficient $B$ by matching it to Eqs.~\eqref{master_bc1} and~\eqref{master_bc2} at $x=x_s$
\begin{equation}
B = \frac{\mathcal{R}}{A_{\mathrm{in}} e^{-2i\omega x_s}-\mathcal{R}A_{\mathrm{out}}} .
\end{equation}
In the derivation, we have made use of the assumption $x_s \gg 1$, so that the asymptotic expressions for $f_\pm$ are valid in the region $x\sim x_s$.

\begin{acknowledgments}
We thank anonymous referees for constructive suggestions, based on which the manuscript has been improved.
This work is supported by National Key R\&D Program of China (No. 2020YFC2201400)
and National Natural Science Foundation of China (NNSFC) under contract Nos. 11805166, 11775036, and 11675139.
We also gratefully acknowledge the financial support from
Funda\c{c}\~ao de Amparo \`a Pesquisa do Estado de S\~ao Paulo (FAPESP),
Funda\c{c}\~ao de Amparo \`a Pesquisa do Estado do Rio de Janeiro (FAPERJ),
Conselho Nacional de Desenvolvimento Cient\'{\i}fico e Tecnol\'ogico (CNPq),
Coordena\c{c}\~ao de Aperfei\c{c}oamento de Pessoal de N\'ivel Superior (CAPES),
A part of this work was developed under the project Institutos Nacionais de Ciências e Tecnologia - Física Nuclear e Aplicações (INCT/FNA) Proc. No. 464898/2014-5.
This research is also supported by the Center for Scientific Computing (NCC/GridUNESP) of the S\~ao Paulo State University (UNESP).
\end{acknowledgments}

\bibliographystyle{JHEP}
\bibliography{references_echoes,references_qian}

\end{document}